\documentclass[12pt,aps,prd,preprint, tightenlines,superscriptaddress,amsmath,amssymb,nofootinbib]{revtex4-1} 

\usepackage[letterpaper,dvips,hmargin={1in,1in},vmargin={1in,1in}]{geometry}

\RequirePackage[colorlinks=true
,urlcolor=blue
,anchorcolor=blue
,citecolor=blue
,filecolor=blue
,linkcolor=blue
,menucolor=blue
,linktocpage=true
,pdfproducer=medialab
,pdfa=true
]{hyperref}

\allowdisplaybreaks

\usepackage{amsmath,amssymb,amsthm,amsfonts}
\usepackage{graphicx}
\usepackage{xspace}
\usepackage{subfigure}
\usepackage{dcolumn}	% Align table columns on decimal point
\usepackage{hyperref}      	% hypertext links %%ARXIV
\usepackage{bm}		% bold math
\usepackage{epsfig}
\usepackage{epstopdf}
\usepackage{setspace}
\usepackage[usenames, dvipsnames]{color}
\usepackage{slashed}
\usepackage{comment}
\usepackage{enumitem}
\usepackage{siunitx}
\usepackage{array,multirow}
\usepackage{feynmp}
\usepackage{datetime}
\usepackage{enumitem}
\usepackage{titlesec, titletoc}
\titleformat*{\section}{\boldmath\bfseries}
\titleformat*{\subsection}{\boldmath\bfseries}
\setlist[description]{leftmargin=0.4cm}

\makeatletter
\def\endfmffile{%
	\fmfcmd{\p@rcent\space the end.^^J%
		end.^^J%
		endinput;}%
	\if@fmfio
	\immediate\closeout\@outfmf
	\fi
	\ifnum\pdfshellescape>\z@
	\immediate\write18{mpost \thefmffile}%
	\fi}
\makeatother
\newcommand{\PRE}[1]{{#1}} % Use if preprint style

\newcommand{\be}{\begin{equation}\begin{aligned}}
\newcommand{\ee}{\end{aligned}\end{equation}}
\newcommand{\beq}{\begin{equation}}
\newcommand{\eeq}{\end{equation}}
\newcommand{\beqa}{\begin{eqnarray}}
\newcommand{\eeqa}{\end{eqnarray}}

\newcommand{\ev}{\text{eV}}

\newcommand{\mev}{\text{MeV}}
\newcommand{\gev}{\text{GeV}}

\newcommand{\cm}{\text{cm}}
\newcommand{\m}{\text{m}}

\newcommand{\s}{\text{s}}

\renewcommand{\eqref}[1]{Eq.~(\ref{eq:#1})}

\newcommand{\secref}[1]{Sec.~\ref{sec:#1}}

\newcommand{\figref}[1]{Fig.~\ref{fig:#1}}

\newcommand{\slas}[1]{\! \not{\! \! #1}}

\RequirePackage[normalem]{ulem}

\def\ol#1{\overline{#1}}

\def\nl{\nonumber\\}

\begin{document}
 
%The following line prevents footnotes from touching the bottom of the page - it essentially fixes a bug in revtex
\count\footins = 1000
 
\preprint{UCI-TR-2021-12}
\PRE{\vspace*{0.3in} }

\title{ 
{\large Gamma Factory Searches for \\
Extremely Weakly-Interacting Particles } 
\PRE{\vspace*{0.4in} }
}

\author{Sreemanti Chakraborti}
\email{sreemanti@iitg.ac.in}
\affiliation{Department of Physics, Indian Institute of Technology Guwahati, Assam 781039, India
\PRE{\vspace*{0.1in}}
}

\author{Jonathan~L.~Feng}
\email{jlf@uci.edu}
\affiliation{Department of Physics and Astronomy,  University of California, Irvine, CA 92697-4575, USA
\PRE{\vspace*{0.1in}}
}

\author{James~K.~Koga}
\email{koga.james@qst.go.jp}
\affiliation{Kansai Photon Science Institute, National Institutes for Quantum and Radiological Science and Technology, Umemidai 8-1-7, Kizugawa, Kyoto 619-0215, Japan
\PRE{\vspace*{0.3in}}
}

\author{Mauro Valli\PRE{\vspace*{0.2in}}
}
\email{mvalli@uci.edu}
\affiliation{Department of Physics and Astronomy,  University of California, Irvine, CA 92697-4575, USA
\PRE{\vspace*{0.1in}}
}

\begin{abstract}
\PRE{\vspace*{0.2in}}
The Gamma Factory is a proposal to back-scatter laser photons off a beam of partially-stripped ions at the LHC, producing a beam of $\sim 10~\mev$ to $1~\gev$ photons with intensities of $10^{16}$ to $10^{18}~\text{s}^{-1}$.  This implies $\sim 10^{23}$ to $10^{25}$ photons on target per year, many orders of magnitude greater than existing accelerator light sources and also far greater than all current and planned electron and proton fixed target experiments.  We determine the Gamma Factory's discovery potential through ``dark Compton scattering,'' $\gamma e \to e X$, where $X$ is a new, weakly-interacting particle.  For dark photons and other new gauge bosons with masses in the 1~to~100 MeV range, the Gamma Factory has the potential to discover extremely weakly-interacting particles with just a few hours of data and will probe couplings as low as $\sim 10^{-9}$ with a year of running. The Gamma Factory therefore may probe couplings lower than all other terrestrial experiments and is highly complementary to astrophysical probes.  We outline the requirements of an experiment to realize this potential and determine the sensitivity reach for various experimental configurations. 
\end{abstract}

%\pacs{}

%\pagenumbering{gobble}
\maketitle
%\thispagestyle{empty}

%\renewcommand{\baselinestretch}{0.95}\normalsize
%\tableofcontents
%\renewcommand{\baselinestretch}{1.0}\normalsize

%\pagenumbering{arabic}

\clearpage

%%%%%%%%%%%%%%%%%%%%%%%%%%%%%%%%%%%%%%%%%%%
\section{Introduction}
\label{sec:introduction}
%%%%%%%%%%%%%%%%%%%%%%%%%%%%%%%%%%%%%%%%%%%

The search for new light and weakly-interacting particles is currently an area of great interest~\cite{Battaglieri:2017aum,Beacham:2019nyx}.  If new particles have masses in the MeV to GeV range, like most of the known particles, they cannot be coupled to the known particles with ${\cal O}(1)$ couplings. However, loop-suppressed interactions with Standard Model (SM) particles are expected in theories with a dark sector~\cite{Holdom:1985ag}, and the requirement that such dark sectors contain dark matter particles with the desired thermal relic density also motivates such small couplings~\cite{Boehm:2003hm,Feng:2008ya}.  In fact, frameworks have been identified in which the couplings are first generated by anywhere from 1-loop to 6-loop interactions, resulting in couplings in the broad range of $\varepsilon \sim 10^{-3}$ to $10^{-13}$~\cite{Gherghetta:2019coi}.  Clearly the existence of such particles is an open experimental question, and novel searches for such particles should be explored, particularly if they exploit existing facilities (see, e.g., Refs.~\cite{Bjorken:2009mm,Andreas:2012mt}).

The Gamma Factory (GF) is such an initiative, which exploits the Large Hadron Collider (LHC)~\cite{Krasny:2015ffb,Budker:2020zer,Dutheil:2020ekk}. In this proposal, laser light with energy $E_{\text{laser}} \sim 10~\ev$ is back-scattered off partially-stripped ions that are accelerated in the LHC to Lorentz factors $\gamma \sim 200$ to 3000. Using the same principle that governs radar guns, the laser light is Doppler shifted twice to energies
\begin{equation}
E_{\text{GF}} 
= E_{\text{laser}}
\left( \sqrt{\frac{1+v/c}{1-v/c}} \ \right)^2
\approx 4 \gamma^2 E_{\text{laser}}
\sim 10~\mev - 1~\gev \ .
\end{equation}
These energies are well-matched to the MeV to GeV mass range for new, weakly-interacting particles.  Just as remarkable, the expected intensities of $\Phi_{\text{GF}} \sim 10^{16}$ to $10^{18}~\s^{-1}$ are far greater than any other existing or proposed accelerator light source, and the resulting number of GF photons per year, $N_{\text{GF}} \sim 10^{23}$ to $10^{25}$, is significantly greater than the protons on target and electrons on target of all fixed target experiments used to search for new MeV to GeV particles to date. The GF, then, has the potential to explore models with light, weakly-interacting particles in regions of parameter space inaccessible to other experiments.

In this paper, we determine the GF's discovery potential for a variety of new, weakly-interacting particles $X$ produced through dark Compton scattering, $\gamma e \to e X$.  Dark Compton scattering has been considered previously for existing photon beam facilities, which have been shown to provide new sensitivity in regions of parameter space with relatively large couplings $\varepsilon \sim 10^{-5}$ to $10^{-3}$~\cite{Chakrabarty:2019kdd}.  Here we focus on the GF's potential and consider dark photons, ``anomaly-free'' ($B-L$, $L_e-L_\mu$, $L_e - L_\tau$) gauge bosons, dark Higgs bosons, and dark pseudoscalars.  For the last two cases, where couplings are Yukawa-suppressed, dark Compton scattering is not promising; nuclear scattering may be more sensitive, but we will not consider this here.  However, in all of the gauge boson cases, we find that dark Compton scattering at the GF has significant discovery prospects, probing regions of parameter space with masses $m_X \sim 1$ to 100 MeV and couplings $\varepsilon \sim 10^{-9}$ to $10^{-4}$, where the low-$\varepsilon$ part of the range extends to values far lower than all other terrestrial experiments.  The GF is therefore complementary to other ongoing and proposed experiments that make use of the LHC to search for weakly-interacting particles~\cite{Mitsou:2020hmt,Ball:2021qrn,Abreu:2020ddv,Ariga:2018pin,Alpigiani:2018fgd,Aielli:2019ivi,Ahdida:2750060,Batell:2021blf}, and our results provide a significant new physics case for the GF, supplementing existing SM and beyond the SM motivations~\cite{Flambaum:2020bqi,Budker:2020zer,Wojtsekhowski:2021xlh,GFMeeting}.

%%%%%%%%%%%%%%%%%%%%%%%%%%%%%%%%%%%%%%%%%%%
\section{A Fixed Target Experiment}
\label{sec:experiment}
%%%%%%%%%%%%%%%%%%%%%%%%%%%%%%%%%%%%%%%%%%%

\begin{figure}[tb]
\begin{center}
\includegraphics*[width=0.99\textwidth]{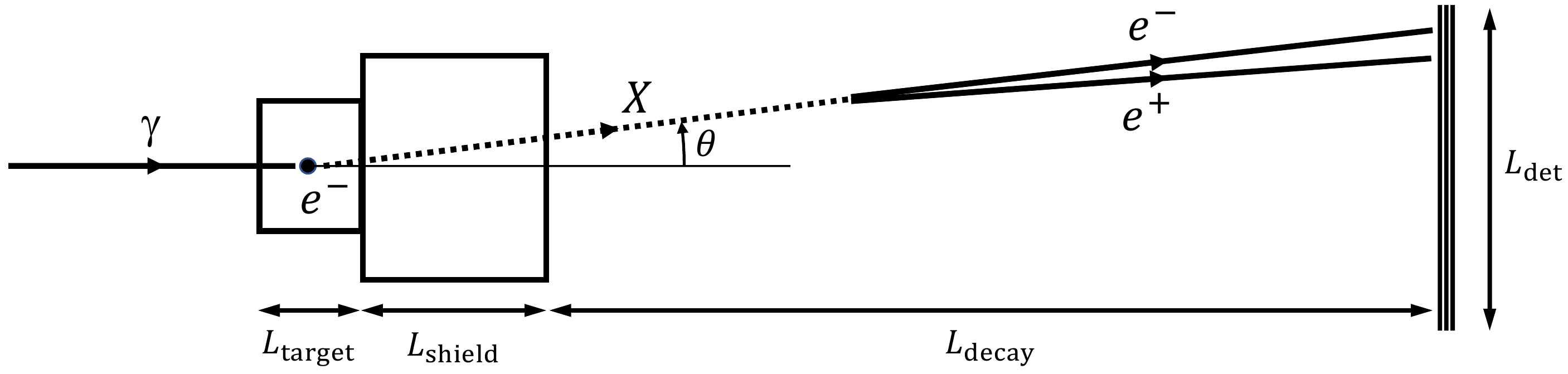}
\end{center}
\vspace*{-0.2in}
\caption{{\bf Experiment layout.} {The experiment consists of a (graphite) target with thickness $L_{\text{target}} = 1~\m$, followed by a (lead) shield with thickness $L_{\text{shield}} = 2~\m$, an open air decay region with length $L_{\text{decay}}$, and a tracking detector, centered on the beam axis, which we take to be a circular disk with diameter $L_{\text{det}}$.  The GF photon beam enters from the left and produces an $X$ particle through dark Compton scattering $\gamma e \to e X$.  The $X$ particle is produced with an angle $\theta$ relative to the GF beamline and decays to an $e^+e^-$ pair, which is detected in the tracking detector.} 
}
\label{fig:experimentlayout}
\end{figure}

The fixed target experiment we propose is simple, compact, and not particularly remarkable; it is shown schematically in \figref{experimentlayout}.  A GF photon beam collides with a target material, producing new particles $X$ through dark Compton scattering $\gamma e \to e X$.  The target is followed immediately by a shield, a large block of matter that stops all SM particles.  The $X$ particles are extremely weakly interacting, however, and so they may pass through the shield and then decay to $e^+ e^-$ pairs, which may be detected in a particle detector.  The detection of coincident $e^+$ and $e^-$ particles that point back to the target provides a striking signal of the production of a new fundamental particle. 

In this section, we discuss the SM background and the required materials and thickness of the target and shield.  We also discuss, in general, the signal rate and its dependence on the $X$ production cross section and decay width.  In the following sections, we will consider specific candidate $X$ particles and determine the sensitivity reach for each of these particles, as well as its dependence on the length of the decay volume $L_{\text{decay}}$ and the transverse size of the detector $L_{\text{det}}$.  

As discussed in \secref{introduction}, the GF will produce a beam of $\sim 10~\mev - \gev$ photons at intensities that are many orders of magnitude beyond current accelerator light sources.  Taking the photon intensity to be $\Phi_{\text{GF}}=10^{17}~\s^{-1}$~\cite{Krasny:2015ffb,Budker:2020zer} at $200~\mev$
and assuming that the back-scattered photon power is fixed by the radio frequency power~\cite{Krasny:2015ffb} resulting in the flux being inversely proportional to the photon energy (see, for example, Eq.~(10) of Ref.~\cite{PhysRevC.103.054603}), we consider three sets of parameters:
\begin{eqnarray}
E_\gamma &=& 20~\mev , \ \ \, \Phi_{\text{GF}} = 10^{18}~\s^{-1}, \ \  N_{\text{GF}} = 3 \times 10^{25} \nonumber \\
E_\gamma &=& 200~\mev , \ \Phi_{\text{GF}} = 10^{17}~\s^{-1}, \ \  N_{\text{GF}} = 3 \times 10^{24}  \nonumber \\
E_\gamma &=& 1.6~\gev , \ \ \Phi_{\text{GF}} = 10^{16}~\s^{-1}, \ \  N_{\text{GF}} = 3 \times 10^{23}  \ ,
\label{eq:GFparameters}
\end{eqnarray}
where the lowest photon energy is based on a longer laser wavelength or lower ion energy, the highest photon energy would be possible with the HE-LHC project~\cite{Zimmermann:2315725, Balkin:2021jdr}, and, in each case, $N_{\text{GF}}$ is simply the number of photons produced in a full year at the corresponding intensity. The photon energies of \eqref{GFparameters} are maximal energies, and the energy distribution may be quite broad; see, e.g., Refs.~\cite{GFMeetingPlaczek,Curatolo:2018lX}.  In detail, however, the distribution depends on the particular atomic transition being used~\cite{atoms8020012}.  To highlight the dependence of our results on the new physics scenarios being probed and minimize the dependence on particular realizations of the GF, we will assume a monoenergetic photon beam with the energies given in \eqref{GFparameters} in determining sensitivity reaches.  The actual sensitivities will be degraded by the energy spread, but this effect will be small away from threshold, and even for $X$ masses near threshold, the degradation will not greatly compromise the discovery prospects of the GF.  For example, if the effective GF intensity is reduced by a factor of 10, given the strong $\varepsilon^4$ dependence of the event rates (see \eqref{Ns}), the reach in $\varepsilon$ will only be reduced by a factor of 1.8.  As we will see, even with such a reduction, the GF's sensitivity reaches extend far beyond existing constraints.   Of course, once the GF is precisely defined, the effect of beam energy spread should be included in a more refined analysis.

These GF photons can then produce $X$ particles through dark Compton scattering in a target material with cross section $\sigma_X \equiv \sigma (\gamma e \to e X)$.  This competes with the far stronger SM processes, which, at these photon energies, are dominated by pair production in the target's nuclear electromagnetic field, with a small component from SM Compton scattering~\cite{Zyla:2020zbs}.  The probability of producing an $X$ particle is
\begin{equation}
P_{\text{prod}} = \frac{Z \sigma_X}{\sigma_{\text{SM}}} \ ,
\label{eq:productionprobability}
\end{equation}
where $Z$ is the number of electrons per target atom, and $\sigma_{\text{SM}}$ is the SM cross section per target atom.  We neglect secondary production of $X$ particles from subsequent processes.  Our analysis is therefore conservative, but these additional sources of $X$ particles are unlikely to enhance significantly the sensitivity reaches we derive.  

Clearly the signal rate is optimized for target materials with low $\sigma_{\text{SM}} / Z$.  Since $\sigma_{\text{SM}}$ is very roughly proportional to $Z^2$, this is minimized for low-$Z$ materials.  For H, Be, and C and the photon energies of interest, the SM cross sections are~\cite{PhotonDatabase} 
\begin{eqnarray}
&& \sigma^{\text{H}}_{\text{SM}}/Z 
= \text{36, 19, 20 mb} \text{ for }
E_{\gamma} = \text{20, 200, 1600 MeV} \\
%&& \sigma^{\text{He}}_{\text{SM}}/Z 
%= \text{39, 25, and 27 mb} \text{ for }
%E_{\gamma} = \text{20, 200, and 1600 MeV} \\
&& \sigma^{\text{Be}}_{\text{SM}}/Z 
= \text{46, 38, 42 mb} \text{ for }
E_{\gamma} = \text{20, 200, 1600 MeV} \\
&& \sigma^{\text{C}}_{\text{SM}}/Z 
= \text{52, 51, 58 mb} \text{ for }
E_{\gamma} = \text{20, 200, 1600 MeV.} 
\end{eqnarray}
For the photons to interact in the target, the target thickness should be a few mean free paths.  At these photon energies, the mean free path is approximately 10 m in liquid hydrogen, 50 cm in beryllium, and 30 cm in graphite~\cite{Zyla:2020zbs,AttenuationCoefficientsDatabase}.  To choose a concrete and practical example for the rest of this analysis, we will assume a graphite target of thickness $L_{\text{target}} = 1~\m$. As we will see, $L_{\text{target}} \ll L_{\text{decay}}$ in the parameter regions of greatest interest, and so for simplicity, we assume that $X$ particles are created with the probability given in \eqref{productionprobability}  with a production point uniformly distributed within $L_{\text{target}}$.

For a background-free experiment, it is ideal, although not necessarily required, for the shield to stop all particles produced by the GF photon beam.  A high-$Z$ material is best, and lead (Pb) is an obvious choice.  The mean free path in Pb for photons with energy $E_{\gamma} \sim 20 - 1600~\mev$ is $\lambda \sim 1-2~\cm$~\cite{Zyla:2020zbs}.  Given an initial number of photons $N_0$, the number remaining after traversing a thickness $L_{\text{shield}}$ of Pb is therefore $N = N_0 e^{-L_{\text{shield}}/\lambda}$.  Thus, even with an initial number of photons $N_0 = 10^{26}$, corresponding to several years of GF running, the number of photons can be reduced to negligible levels for a shield of thickness  $L_{\text{shield}} \sim 60 \lambda \sim 0.6 - 1.2~\m$. We therefore expect that a 2 m thick Pb shield will be sufficient to remove the SM background.\footnote{Depending on the GF setup, photon-nucleus scattering could be a source of muon pair production. However, muons are minimum-ionizing particles that should lose energy roughly as $\sim 2$ MeV/cm. Hence, they are expected to stop in a few meters of the shield here proposed.} The approximate power of the very high photon flux on the target will be $(200\mbox{ MeV})( 1.6022\times 10^{-19}\mbox{ J/eV})(10^{17}\mbox{ s}^{-1})\sim 3 \mbox{ MW}$. This is comparable to the average beam power of 18 MW for the 250 GeV ILC beam dumps~\cite{SATYAMURTHY201267} and 5.3 MW for the 125 GeV ILC beam dumps~\cite{Morikawa:2018}. In addition, the photon beams are narrowly collimated and cannot be spread out to reduce the energy density by magnets, as shown for the photon-photon collider configuration of the ILC with 10--15 MW of power~\cite{Shekhtman_2014}. Therefore, detailed design of cooling systems for the target will be required (see Refs.~\cite{SATYAMURTHY201267,Shekhtman_2014}).

Finally, we must determine the decay volume length $L_{\text{decay}}$ and detector size $L_{\text{det}}$.  As we will see, for all models considered, in the region of parameter space that can be probed for the first time at the GF, the $X$ decay length $d_X = \gamma_X v_X c \tau_X$ is far greater than any reasonable $L_{\text{decay}}$.  The probability of decay in the decay volume is therefore
\begin{equation}
P_{\text{decay}} 
= e^{-(L_{\text{target}}+L_{\text{shield}})/d_X}
- e^{-(L_{\text{target}}+L_{\text{shield}}+L_{\text{decay}})/d_X}
\approx \frac{L_{\text{decay}}}{d_X} \ .
\label{eq:decay}
\end{equation}
The number of signal events scales linearly with $L_{\text{decay}}$, but larger $L_{\text{decay}}$ requires a detector with larger $L_{\text{det}}$ to capture the produced $e^+ e^-$ pairs.  We will explore how the sensitivity depends on $L_{\text{decay}}$ and $L_{\text{det}}$ in the following sections, but as a preview of these results, we will find that parameters $L_{\text{decay}} \sim 10~\m$ and $L_{\text{det}} \sim 1~\m$ will be sufficient to probe large swaths of new parameter space.

%%%%%%%%%%%%%%%%%%%%%%%%%%%%%%%%%%%%%%%%%%%
\section{Dark Photons}
\label{sec:darkphotons}
%%%%%%%%%%%%%%%%%%%%%%%%%%%%%%%%%%%%%%%%%%%

We first consider the case where the new, weakly-interacting particle is the dark photon $A'$~\cite{Holdom:1985ag,Fabbrichesi:2020wbt,Graham:2021ggy}.  The dark photon's properties are determined by two parameters, its mass $m_{A'}$ and its coupling $\varepsilon$ (in units of $e$), which enter the Lagrangian through
\begin{equation}
\mathcal{L} \supset  \frac{1}{2} m_{A'}^2 A'^2
- \varepsilon \, e \sum_f q_f \bar{f} \slas{A'} f \ ,
\end{equation}
where $q_f$ is the SM electric charge of fermion $f$.  

The cross section for dark Compton scattering $\gamma e \to e A'$ and the angular distribution of the produced dark photons are shown in \figref{darkphotonsigma}.  (See the Appendix for further details.)  The cross section is maximal not far above threshold, then drops for increasing $E_{\gamma}$, but remains within an order of magnitude of the maximum for all GF photon energies. The angular distribution of the produced dark photons is also highly peaked in the forward direction. This is clearly true at threshold, since there is no excess energy to support components of the $A'$ momentum transverse to the beam, but we see that it is even true for light dark photons when the beam energy is far above threshold, at least for the beam energy shown. 

\begin{figure}[tb]
  \centering
   \includegraphics[width=0.47\linewidth]{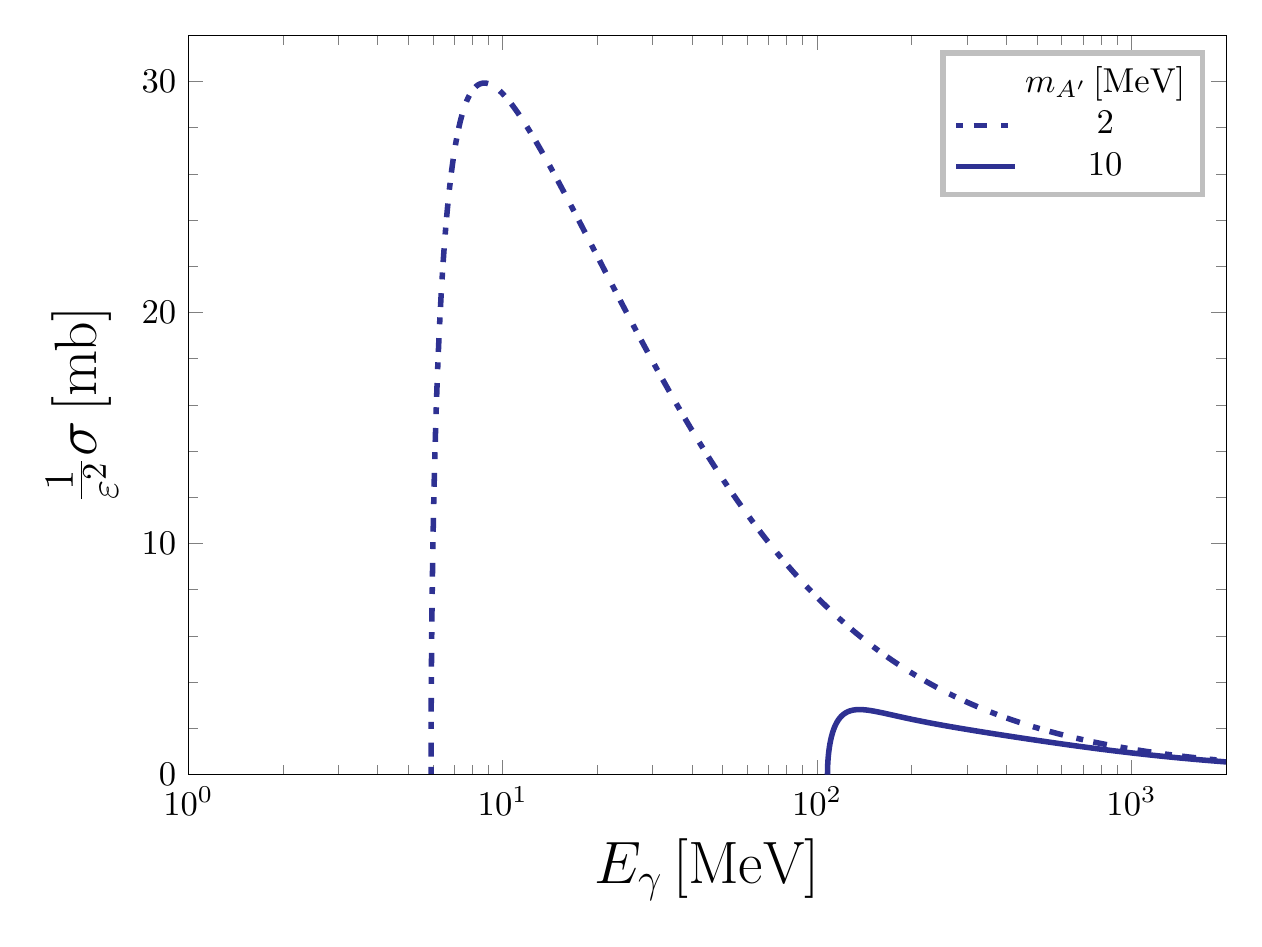} 
  \hfill
   \includegraphics[width=0.50\linewidth]{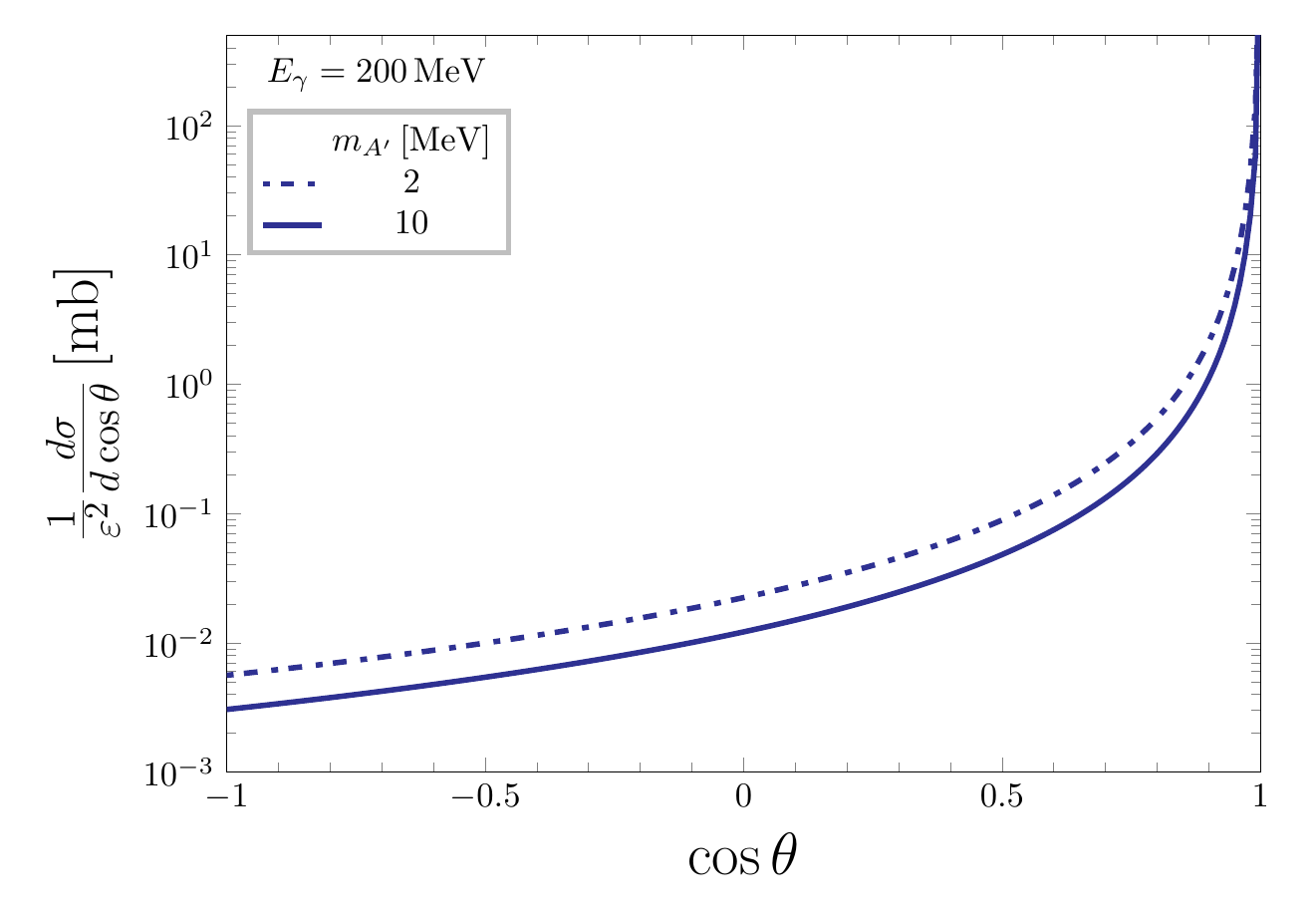}
  \caption{{\bf Left}: { Dark photon production cross section $\sigma(\gamma e \to e A')$ as a function of incoming photon energy $E_\gamma$ for $m_{A'} = 2, 10~\mev$.} {\bf Right}: { Lab frame angular differential distribution of dark photons $A'$ produced through dark Compton scattering, where $\theta$ is the angle relative to the photon beam line (see \figref{experimentlayout}), $E_{\gamma} = 200~\mev$, and $m_{A'} = 2, 10~\mev$. }}
  \label{fig:darkphotonsigma}
\end{figure}

Once produced, the dark photon dominantly decays to pairs of SM particles, assuming $m_{A'} > 2 m_e$.  For $m_{A'} > 2 m_{\mu}$, decays to muons and a number of hadronic states are possible, but, given the available GF energies of \eqref{GFparameters}, $m_{A'} \alt 40~\mev$, and so only the decay channel ${A'} \to e^+ e^-$ is open.  We assume that there are no non-SM decays.  In this case, the dark photon decay width is
\begin{equation}
\Gamma_{A'} = \Gamma (A' \to e^+ e^-) 
=  \frac{\varepsilon^2 e^2 m_{A'}}{12 \pi} 
\Biggl[ 1 - \left( \frac{2 m_e} {m_{A'}} \right)^{\! \! 2} \ \Biggr]^{1/2} 
\Biggl[ 1 + \frac{2 m_e^2}{m_{A'}^2} \Biggr] \approx \frac{\varepsilon^2 e^2 m_{A'}}{12 \pi} \ ,
\end{equation}
where in the last expression, we have assumed $m_{A'} \gg m_e$.  If the ${A'}$ is produced relativistically, with $v_{A'} \approx 1$ and $\gamma_{A'} \equiv E_{A'} / m_{A'} \gg 1$, its decay length is
\begin{equation}
d_{A'} = \gamma_{A'} v_{A'} c \, \frac{1}{\Gamma_{A'}} 
\approx 8.1 \times 10^5~\m \left[ \frac{10^{-8}}{\varepsilon} \right]^2 
\left[ \frac{E_{A'}}{100~\mev} \right] \left[ \frac{10~\mev}{m_{A'}} \right]^2 \ .
\end{equation}
We see that in the region of parameter space where the GF will probe new parameter space, $d_{A'} \gg L_{\text{decay}}$, as anticipated in \eqref{decay}. The probability of decay within the decay volume is very small, and this must be compensated by producing an extraordinarily large number of dark photons.

To determine the sensitivity reach, for any parameters $(m_{A'}, \varepsilon)$, we simulate dark photon production by dark Compton scattering, including the correct $\cos \theta$ distribution.  In particular, using a Monte Carlo approach, we sample $X$ particle momenta, weighted by the matrix element of the production process.  We then decay the dark photon to $e^+ e^-$ pairs, according to the probability distribution given in \eqref{decay}, with the approximation that the decays are isotropic in the $A'$ rest frame.  
Practically, for a given point in parameter space, i.e., for a fixed pair of $X$ mass and coupling, we randomly extract $10^5$ values of $\cos \theta$ from the inverse of the cumulative distribution function:
\begin{equation}
    \mathcal{P}(\cos \theta) = \frac{\int_{-\cos \theta}^{\cos \theta} \overline{|\mathcal{M}|^2}}{\int_{-1}^{1} \overline{|\mathcal{M}|^2}} \ \in \ [0,1] \, ,
\end{equation}
where $\overline{|\mathcal{M}|^2}$ denotes the spin-averaged matrix element of the dark Compton scattering process. From the  distribution of $\cos \theta$ so obtained, we eventually derive the distribution of the signal events, $\mathcal{P}(N_{S})$. In particular, after checking that the simulated $e^{\pm}$ pairs pass through the detector, we can compute the mean of events $\langle N_{S} \rangle$. If $\langle N_{S} \rangle \geq$ 3 events, we accept the chosen point in parameter space as one within the GF sensitivity. In any other case, we discard it.
A signal event is indeed defined to be an event where both the $e^+$ and the $e^-$ pass through the tracking detector shown in \figref{experimentlayout}.  The coincident detection of two oppositely-charged particles, each pointing back to the target, will be a striking signal, and we will assume zero background.  If the $e^+$ and $e^-$ energies can be measured, for example, by placing the tracker in a strong magnetic field or adding a calorimeter, the invariant mass of the $e^+e^-$ pair can be determined, providing a further kinematic constraint to differentiate signal from background, as well as a measurement of the $A'$ mass.

\begin{figure}[tb]
\begin{center}
\includegraphics*[width=0.8\textwidth]{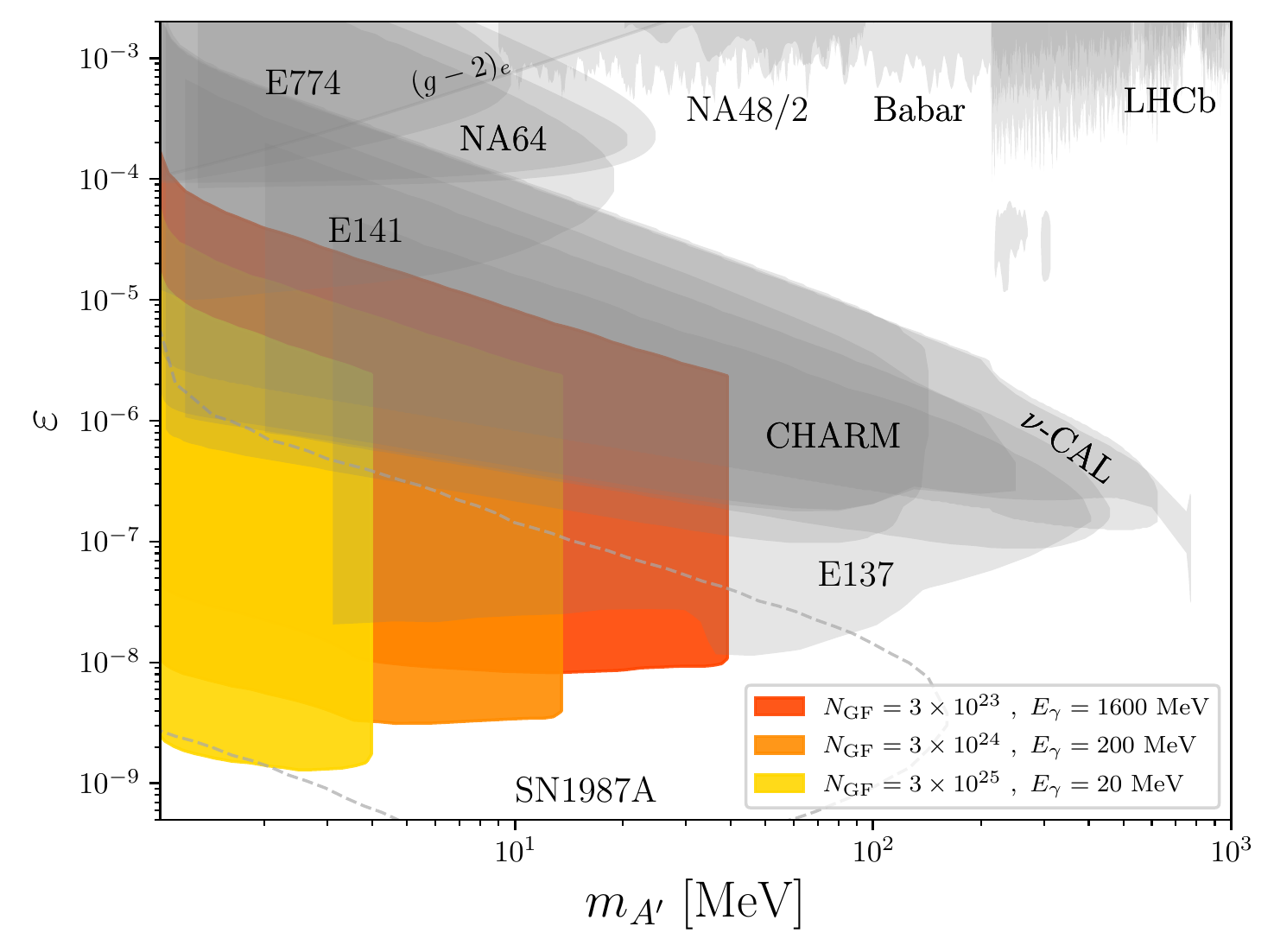}
\hfill
\end{center}
\vspace*{-0.3in}
\caption{{\bf Dark photon sensitivity.} { The sensitivity reach for the three sets of GF parameters $(E_{\gamma}, N_{\text{GF}})$ indicated, each corresponding to a year of running, and detector parameters $L_{\text{decay}} = 12~\m$ and $L_{\text{det}} = 3~\m$.  The contours are for 3 $e^+e^-$ signal events and assume no background. The gray shaded regions are existing bounds from the terrestrial experiments indicated~\cite{Bergsma:1985qz,Gninenko:2012eq,Riordan:1987aw,Bjorken:1988as,Davier:1989wz,Bross:1989mp,Blumlein:1990ay,Blumlein:1991xh,Lees:2014xha,Batley:2015lha,Marsicano:2018krp,NA64:2019imj,Aaij:2019bvg} (for further details, see also~\cite{Ilten:2018crw,Bauer:2018onh}), from $(g-2)_e$~\cite{Endo:2012hp}, and the dashed gray line encloses the region probed by supernova cooling, as determined in Ref.~\cite{Chang:2018rso}.}
\label{fig:darkphotonreach}}
\end{figure}

The sensitivity reach is shown in \figref{darkphotonreach}.  These results may be understood as follows:  The sensitivity regions are bounded at low mass by the requirement that the $e^+e^-$ decay is open ($m_{A'} > 2 m_e$) and at high mass by the requirement that dark Compton scattering $\gamma e \to e X$ is kinematically accessible ($m_{A'} \alt \sqrt{2 m_e E_{\gamma}}$ ).  The regions are further bounded at large $\varepsilon$ by the requirement that the dark photons travel through the target and shield before decaying ($d_{A'} \agt 3~\m$), and at small $\varepsilon$ by the requirement that a sufficient number of dark photons decay in the decay volume.

It is instructive to understand the bound at small $\varepsilon$ by estimating the number of signal events in the limit of long decay lengths. We parametrize $\sigma_X \sim \varepsilon^2 \, (1~\text{mb}) \, (10~\mev/m_{A'})^2$, assume $E_{\gamma} = 200~\mev$ and a typical dark photon energy $E_{A'} \sim 100~\mev$, and let $L_{\text{decay}} = 12~\m$ and $P_{\text{det}} \sim 1$ be the probability that a dark photon that decays in the decay volume is captured in the detector.  The signal event rate is, then, roughly
\begin{eqnarray}
N_S &=& N_{\text{GF}} P_{\text{prod}} P_{\text{decay}} P_{\text{det}} 
\sim  N_{\text{GF}} \frac{Z \sigma_X}{\sigma_{\text{SM}}} \frac{L_{\text{decay}}}{d_{A'}} \nonumber \\
&\sim & N_{\text{GF}} \frac{6 \ \varepsilon^2 \ 1~\text{mb}}{50~\text{mb}} \left[ \frac{10~\mev}{m_{A'}} \right]^2 \frac{12~\m}{6.5\times 10^{5}~\m} \left[ \frac{\varepsilon}{10^{-8}} \right]^2 \left[ \frac{m_{A'}}{10~\mev} \right]^2 \nonumber \\
&=& 3 \, \frac{N_{\text{GF}}}{3 \times 10^{24}} \left[ \frac{\varepsilon}{2.6 \times 10^{-9}} \right]^4 \ .
\label{eq:Ns}
\end{eqnarray}
We see that, provided the beam energy is above threshold, the number of events is approximately independent of $m_{A'}$, but is highly sensitive to $\varepsilon$.  One also expects to probe $\varepsilon$ as low as $10^{-9}$, given the extraordinary number of GF photons on target.  All of these features are confirmed by the simulation results shown in \figref{darkphotonreach}.

The GF probes new parameter space at low values of $\varepsilon$ between $10^{-9}$ and $10^{-7}$.  Such low values are inaccessible to all other terrestrial experiments investigated to date, because the signal rate is suppressed by low production rates and the long $A'$ decay length.  At the GF, however, this suppression is compensated by the extraordinary number of photons on target.  Such low values of $\varepsilon$ are subject to astrophysical constraints, for example, from supernova cooling~\cite{Dent:2012mx,Dreiner:2013mua,Kazanas:2014mca,Rrapaj:2015wgs,Chang:2016ntp,Chang:2018rso,Sung:2019xie}.  However, such constraints are dependent on a number of astrophysical assumptions, which may weaken the constraints or possibly even remove them altogether; see, e.g., Ref.~\cite{Bar:2019ifz}. The GF therefore probes a significant new region of parameter space that cannot be probed by other particle experiments, and it is highly complementary to astrophysical probes.

\begin{figure}[tb]
\begin{center}
\includegraphics*[width=0.495\textwidth]{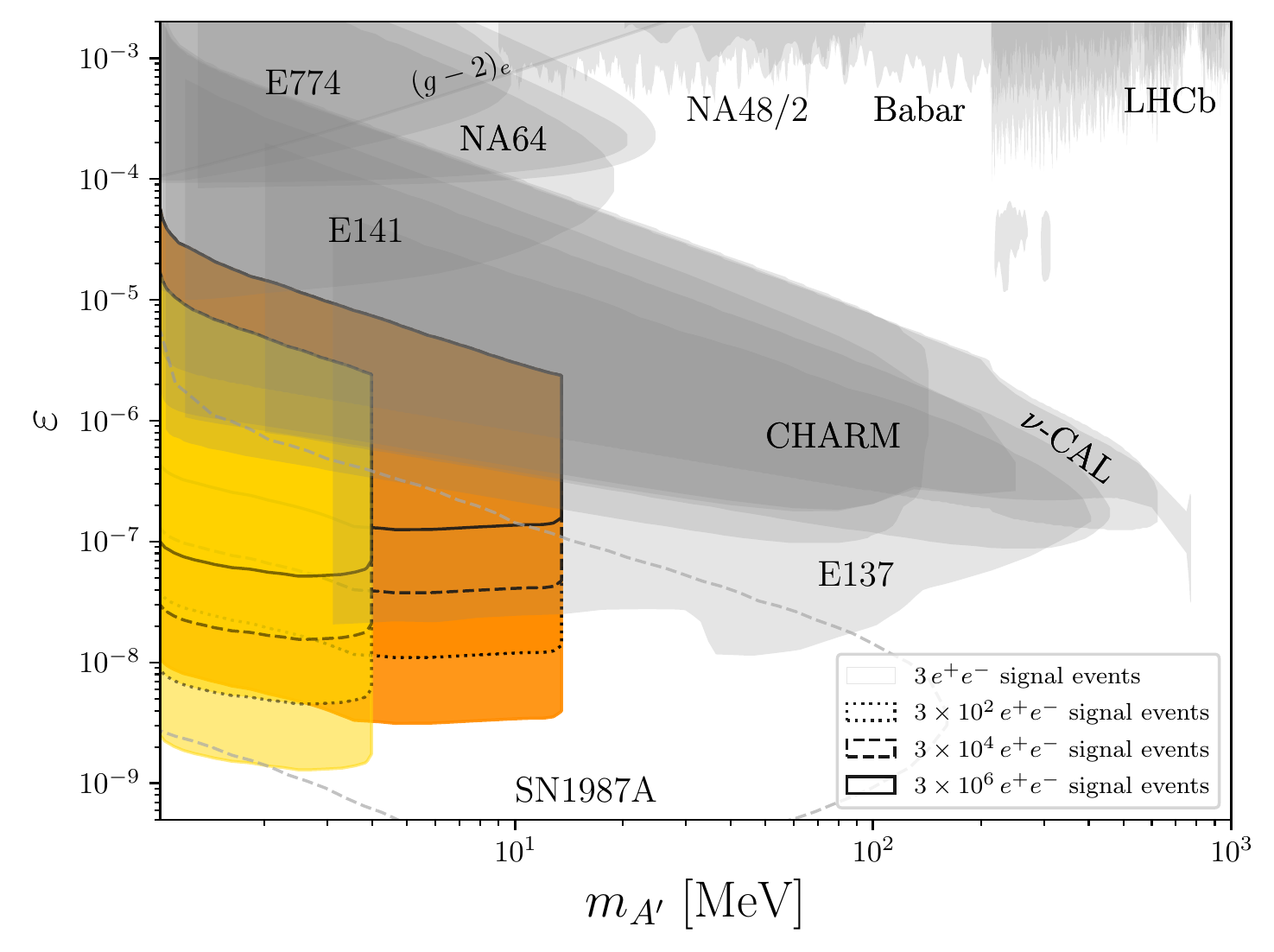}
\hfill 
\includegraphics*[width=0.495\textwidth]{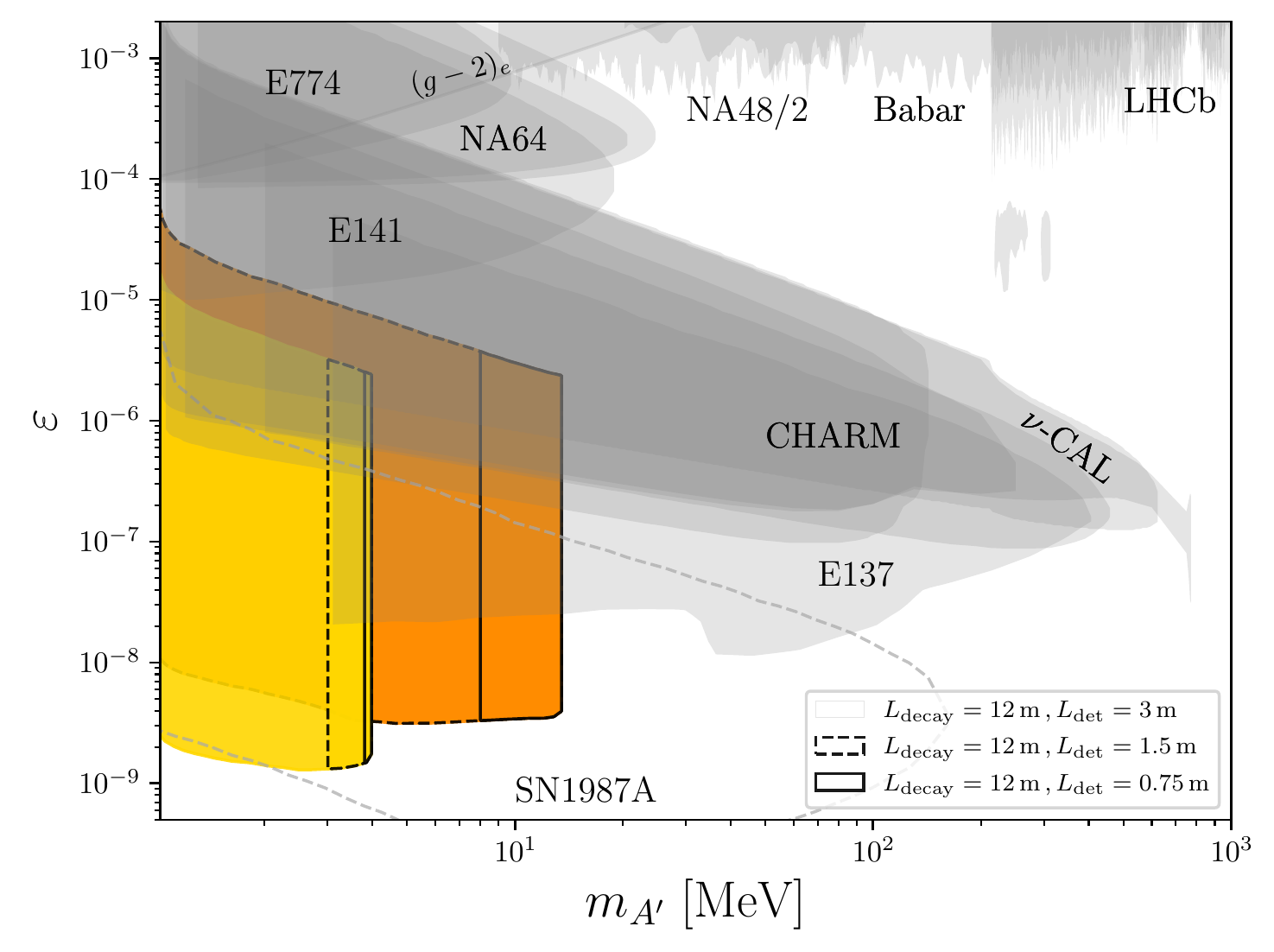}
\end{center}
\vspace*{-0.3in}
\caption{{\bf Left:} { Event rate contours for $(E_{\gamma}, N_{\text{GF}}) = (20~\mev, 3 \times 10^{25})$ (yellow) and $(E_{\gamma}, N_{\text{GF}}) = (200~\mev, 3 \times 10^{24})$ (orange), $L_{\text{decay}} = 12~\m$, and $L_{\text{det}} = 3~\m$.}
{\bf Right:} { The sensitivity reach for $(E_{\gamma}, N_{\text{GF}}) = (20~\mev, 3 \times 10^{25})$ (yellow) and $(E_{\gamma}, N_{\text{GF}}) = (200~\mev, 3 \times 10^{24})$ (orange), $L_{\text{decay}} = 12~\m$, and $L_{\text{det}} = 0.75$, 1.5, and 3 m. The contours are for 3 $e^+e^-$ signal events and assume no background. The gray shaded regions and dashed gray line indicate existing constraints from terrestrial experiments and supernovae, respectively, as in \figref{darkphotonreach}.}
\label{fig:darkphotonreach_variations}}
\end{figure}

In the left panel of \figref{darkphotonreach_variations}, we show signal event rate contours for the GF parameters $(E_{\gamma}, N_{\text{GF}}) = (20~\mev, 3 \times 10^{25})$ (yellow) and $(E_{\gamma}, N_{\text{GF}}) = (200~\mev, 3 \times 10^{24})$ (orange).  Given the strong $\varepsilon$ dependence of \eqref{Ns}, we see that there are uncharted regions of parameter space where as many as $3 \times 10^4$ dark photons could be produced in a year. Assuming a background-free experiment, a dark photon discovery could be achieved with just a few hours of running.  Alternatively, if there is background, one can see that requiring, say, 10 or 100 signal events does not reduce the sensitivity region much, given the dependence of the signal rate on $\varepsilon^4$.

In the right panel of \figref{darkphotonreach_variations}, we show the dependence of the sensitivity reach on the size of the detector $L_{\text{det}}$. For $L_{\text{decay}} = 12~\m$, and $L_{\text{det}} = 3~\m$, the detector is large enough to catch all signal events, and so is effectively infinite in size.  For $L_{\text{det}} = 1.5~\m$ and 0.75 m, however, events may be lost.  This degrades the reach primarily at low $m_{A'}$: for $(E_{\gamma}, N_{\text{GF}}) = (20~\mev, 3 \times 10^{25})$, the low $m_{A'}$ coverage is degraded significantly for $L_{\text{det}} = 1.5~\m$ and almost all coverage is lost for $L_{\text{det}} = 0.75~\m$, while for $(E_{\gamma}, N_{\text{GF}}) = (200~\mev, 3 \times 10^{24})$, the degradation is minimal for $L_{\text{det}} = 1.5~\m$, but again becomes significant for $L_{\text{det}} = 0.75~\m$.  This may be understood as follows: for low masses, there is sufficient energy for the dark photon to be produced with significant transverse momentum, and so one or both of the $e^+$ and $e^-$ particles produced escape detection.  On the other hand, for large $m_{A'}$ near threshold, the dark photons are produced in the direction of the photon beam.  When they decay, the $e^+ e^-$ pairs are produced with some transverse momentum, but this is typically small enough so that no events are lost.  For example, for $m_{A'} = 10~\mev$ and $E_{A'} \sim 100~\mev$, the typical angle of the $e^{\pm}$ relative to the beamline is $m_{A'} / (2 E_{A'}) \sim 0.05$, and so these particles are detected in a detector with size $L_{\text{det}} \sim 0.1 L_{\text{decay}}$.  

\begin{figure}[tb]
\begin{center}
\includegraphics*[width=0.5\textwidth]{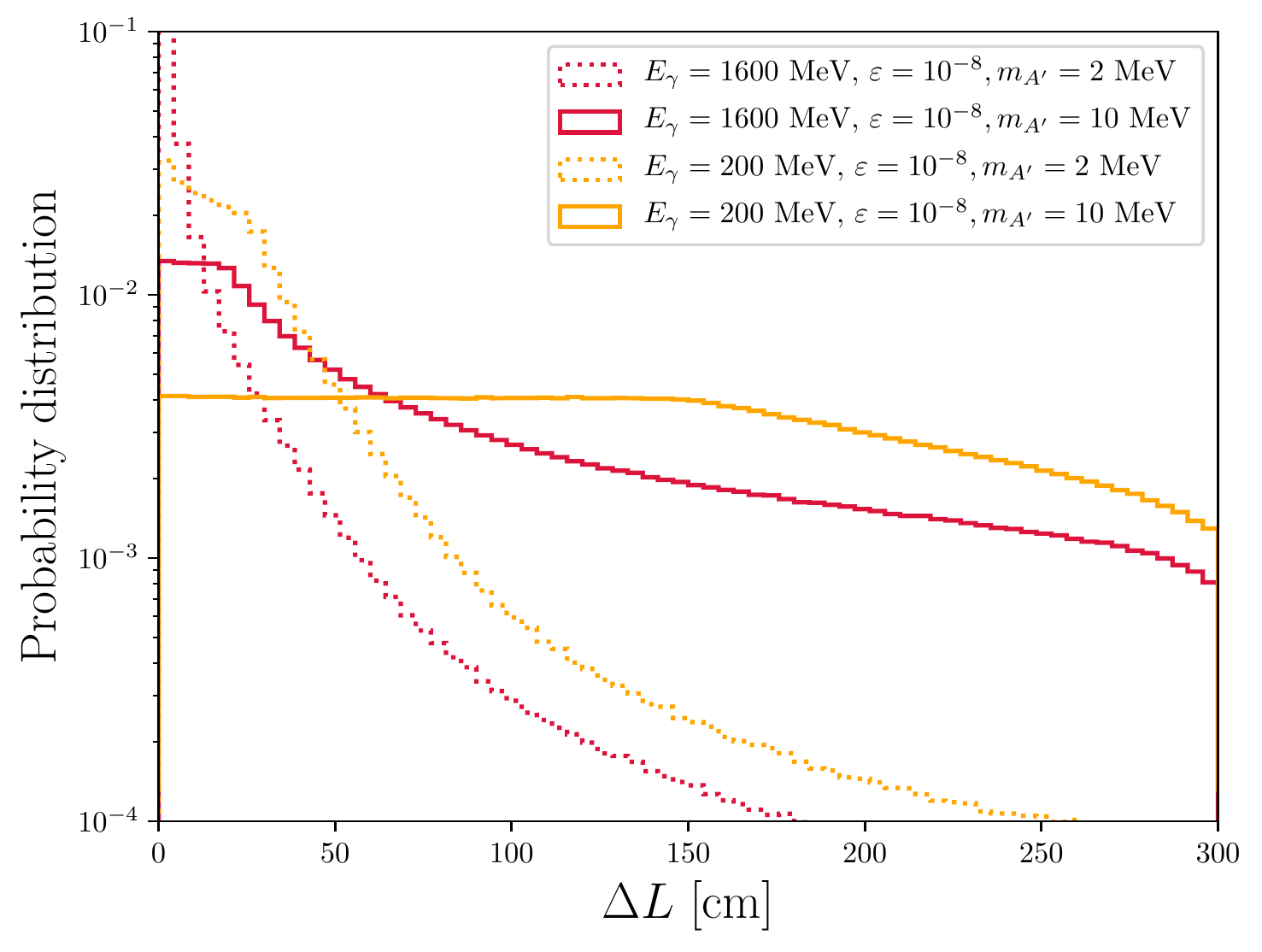}
\end{center}
\vspace*{-0.3in}
\caption{{\bf Separation between \boldmath$e^+$ and \boldmath$e^-$ in signal events.} { For $E_{\gamma}$ and the dark photon parameters $(m_{A'}, \varepsilon)$ indicated, $L_{\text{decay}} = 12~\m$, and $L_{\text{det}} = 3~\m$, we show the distribution of distances between the $e^+$ and $e^-$ when they pass through the detector.} 
\label{fig:darkphotonreach_separation}}
\end{figure}

Finally, in \figref{darkphotonreach_separation}, we show the distribution of distances between the $e^+$ and $e^-$ when they pass through the detector for several representative $E_{\gamma}$ and dark photon parameters.  For $m_{A'} = 10~\mev$, the separations are $\sim 10~\cm - 1~\m$; for $m_{A'} = 2~\mev$, the $e^+$ and $e^-$ are more collimated, as expected, and their separations are reduced to $\sim 1 -10~\cm$. Nevertheless, in all cases shown, the typical separations are large compared to the position resolution of typical trackers, and so the $e^+$ and $e^-$ are easily distinguished in a tracker.  With 2 or more tracking layers, one can also verify that the $e^+$ and $e^-$ are coming from the direction of the GF photon beam.  Although we do not discuss a detailed detector design here, such kinematic constraints can be powerfully exploited to differentiate signal from background.

%%%%%%%%%%%%%%%%%%%%%%%%%%%%%%%%%%%%%%%%%%%
\section{Anomaly-Free Gauge Bosons}
\label{sec:othergaugebosons}
%%%%%%%%%%%%%%%%%%%%%%%%%%%%%%%%%%%%%%%%%%%

The GF also has significant potential to discover other light gauge bosons. We will consider the three cases of gauge bosons that mediate the ``anomaly-free'' U(1) gauge interactions $B-L$, $L_e - L_{\mu}$, and $L_e - L_{\tau}$.\footnote{We do not consider $L_{\mu} - L_{\tau}$ gauge bosons, because their coupling to electrons is generated only at loop-level, and so the GF does not provide a sensitive probe.}  These gauge bosons are included through the additional Lagrangian terms
\begin{equation}
\mathcal{L} \supset  \frac{1}{2} m_{X}^2 X^2
- g_X X^{\mu} j_{\mu}^{X} \ , 
\end{equation}
where $j_{\mu}^X$ is the appropriate current. 

\begin{figure}[tb]
\begin{center}
\includegraphics*[width=0.495\textwidth]{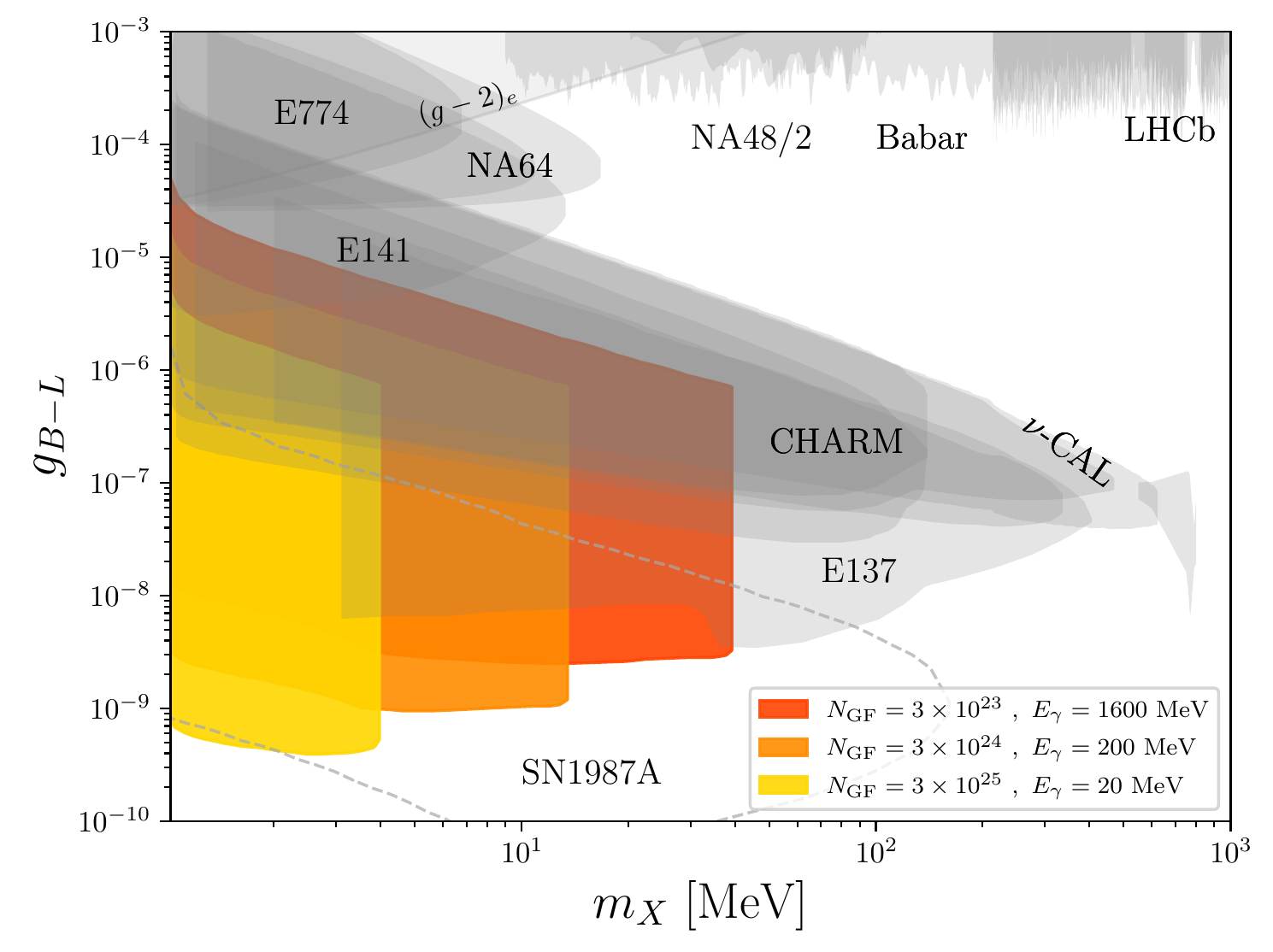}
\includegraphics*[width=0.495\textwidth]{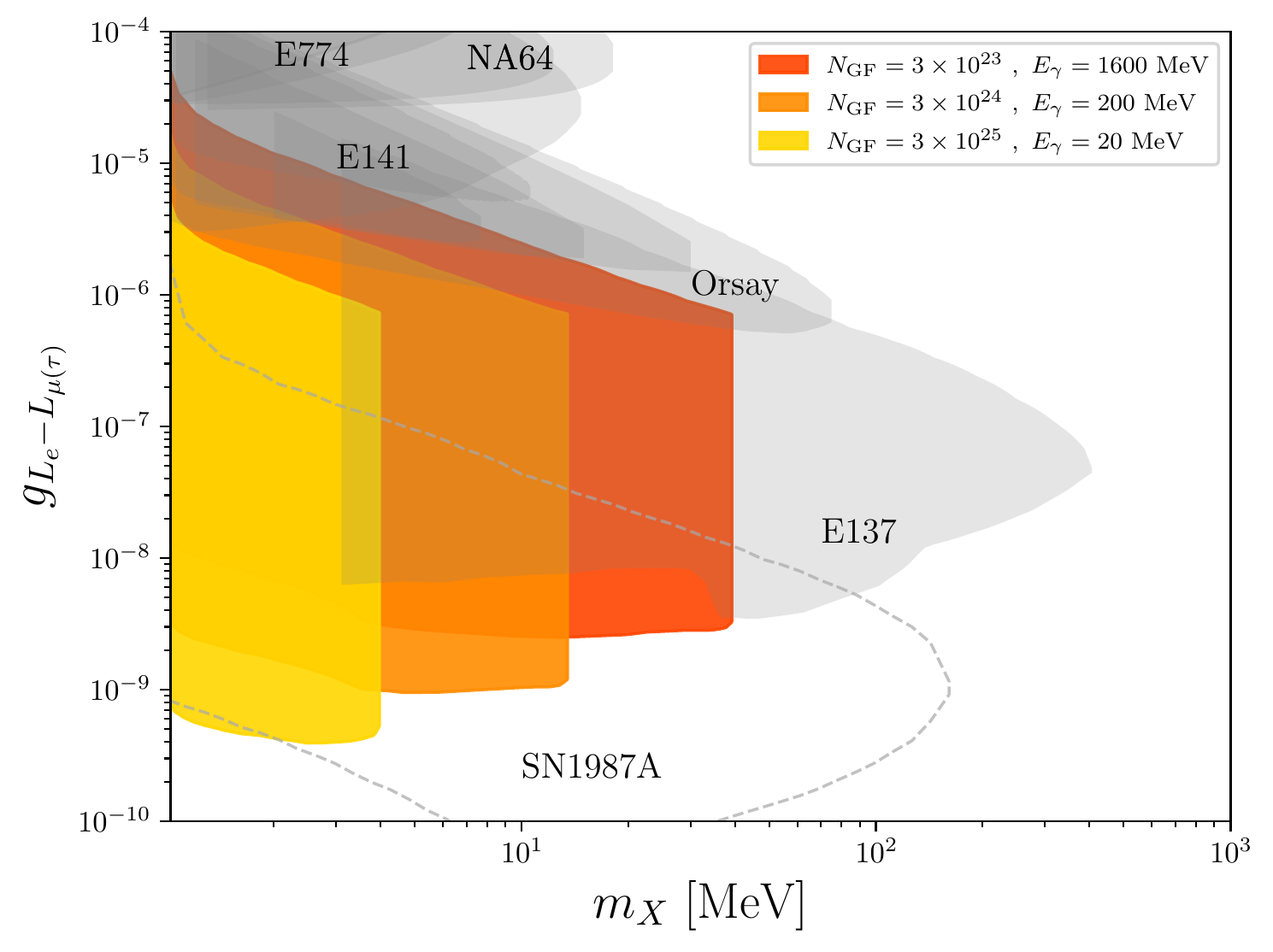}
\end{center}
\vspace*{-0.3in}
\caption{{\bf Sensitivity for anomaly-free gauge bosons.} { The sensitivity reaches for $B-L$ (left) and $L_e - L_{\mu,\tau}$ (right) gauge bosons are shown for the three sets of GF parameters $(E_{\gamma}, N_{\text{GF}})$ indicated and detector parameters $L_{\text{decay}} = 12~\m$ and $L_{\text{det}} = 3~\m$.  The contours are for 3 $e^+e^-$ signal events and assume no background. The gray shaded regions and dashed gray line indicate existing constraints from terrestrial experiments and supernovae, respectively, as in \figref{darkphotonreach}.}
\label{fig:othergaugebosonreach}}
\end{figure}

We simulate the production of these anomaly-free gauge bosons through dark Compton scattering $\gamma e \to e X$, following the same procedure used for dark photons in \secref{darkphotons}.  Unlike in the case of dark photons, in the anomaly-free gauge boson cases, decays to neutrinos are open, reducing the decay lengths, but otherwise the analysis is very similar\footnote{In determining the sensitivity to anomaly-free gauge bosons, we take into account the smaller branching fraction of $X$ into $e^{\pm}$ due to additional decay modes in (massless) neutrinos.  (The branching fraction into $e^{\pm}$ is roughly 2/5 for the case of $B-L$ and 1/2 for $L_{e}-L_{\mu(\tau)}$.) Given the scaling of the signal with the fourth power of the coupling of $X$, the impact of this reduced branching fraction in the estimated sensitivity of \figref{othergaugebosonreach} is almost imperceptible.}. In the parameter space of greatest interest, the results for $L_e - L_{\mu}$ and $L_e - L_{\tau}$ bosons are identical.  The sensitivity reaches for the $B-L$ and $L_e - L_{\mu,\tau}$ cases are shown in \figref{othergaugebosonreach}.

As in the case of dark photons, the GF is able to probe new parameter space for couplings $g_X$ that are far below the reach of all other terrestrial experiments, and the GF's sensitivity is complementary to supernovae probes. 

%%%%%%%%%%%%%%%%%%%%%%%%%%%%%%%%%%%%%%%%%%%
\section{Dark Higgs Bosons and Pseudoscalars}
\label{sec:darkhiggsbosons}
%%%%%%%%%%%%%%%%%%%%%%%%%%%%%%%%%%%%%%%%%%%

For completeness, we consider two spin-0 dark mediator particles (see, e.g., Refs.~\cite{Dolan:2014ska,Winkler:2018qyg}): the dark Higgs boson $\phi$, with Lagrangian terms
\begin{equation}
\mathcal{L} \supset - m_\phi^2 \phi^2 - \sin \alpha \sum_f \frac{m_f}{v} \, \phi \bar{f} f  
% - \lambda \, v \,h \phi \phi 
\ ,
\label{eq:darkHiggseffL}
\end{equation}
and the dark pseudoscalar $a$, with Lagrangian terms
\begin{equation}
\mathcal{L} \supset - m_a^2 a^2 + i g_Y a \sum_f \frac{m_f}{2v} \bar{f} \gamma_5 \, f \ ,
\label{eq:darkpseudoscalareffL}
\end{equation}
where $v \simeq 246~\gev$ is the SM Higgs vacuum expectation value.

The dark Compton scattering production cross sections of spin-0 bosons is detailed in the Appendix.  The cross sections are shown in \figref{sigscal}.  As in the spin-1 cases, the cross sections peak near threshold and then drop as $E_{\gamma}$ increases, but for all GF energies, the cross sections remain within roughly an order of magnitude of their maximum values.

\begin{figure}[tb]
   \centering
  \includegraphics[width=0.49\linewidth]{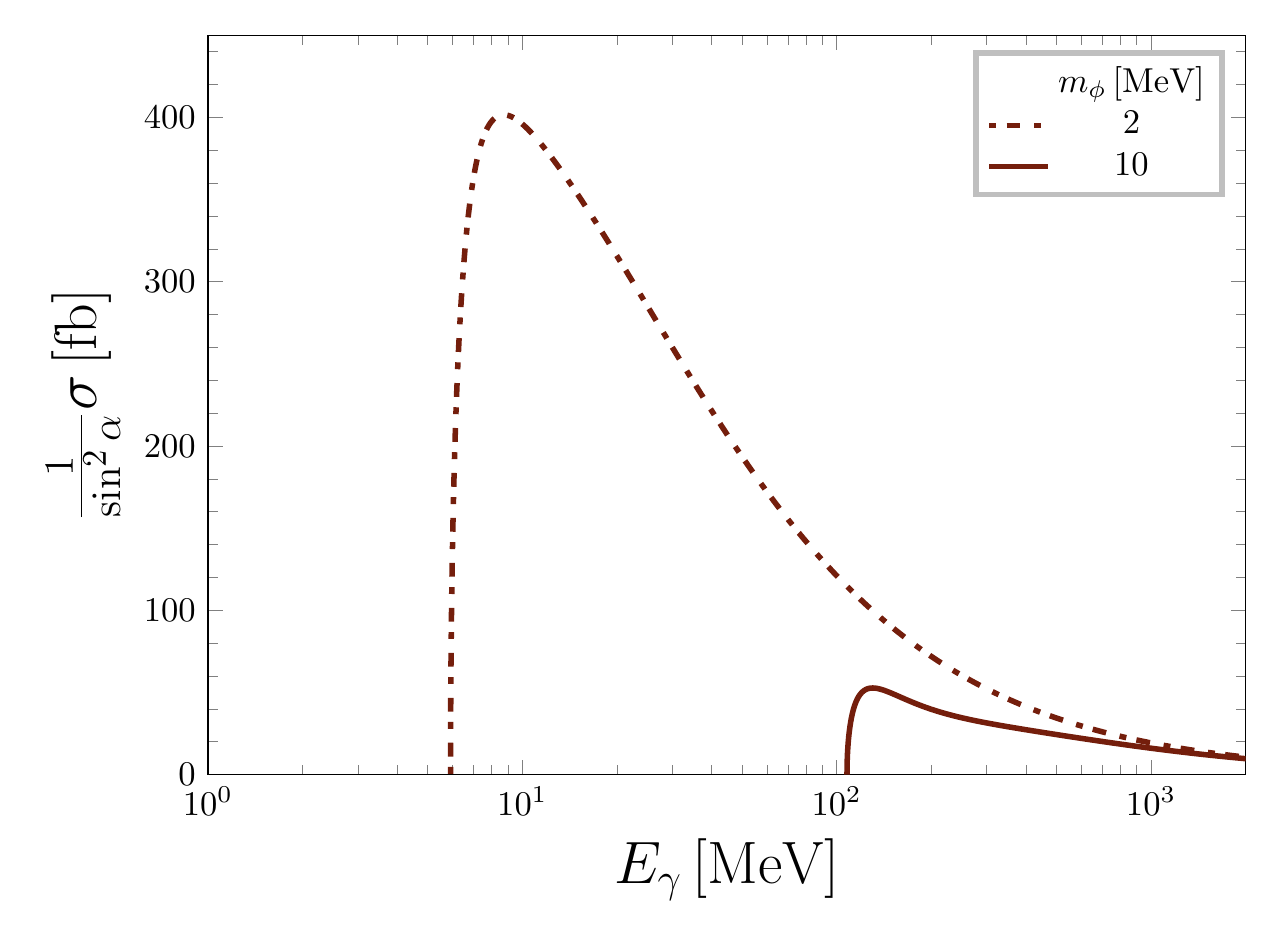}
  \hfill
  \includegraphics[width=0.49\linewidth]{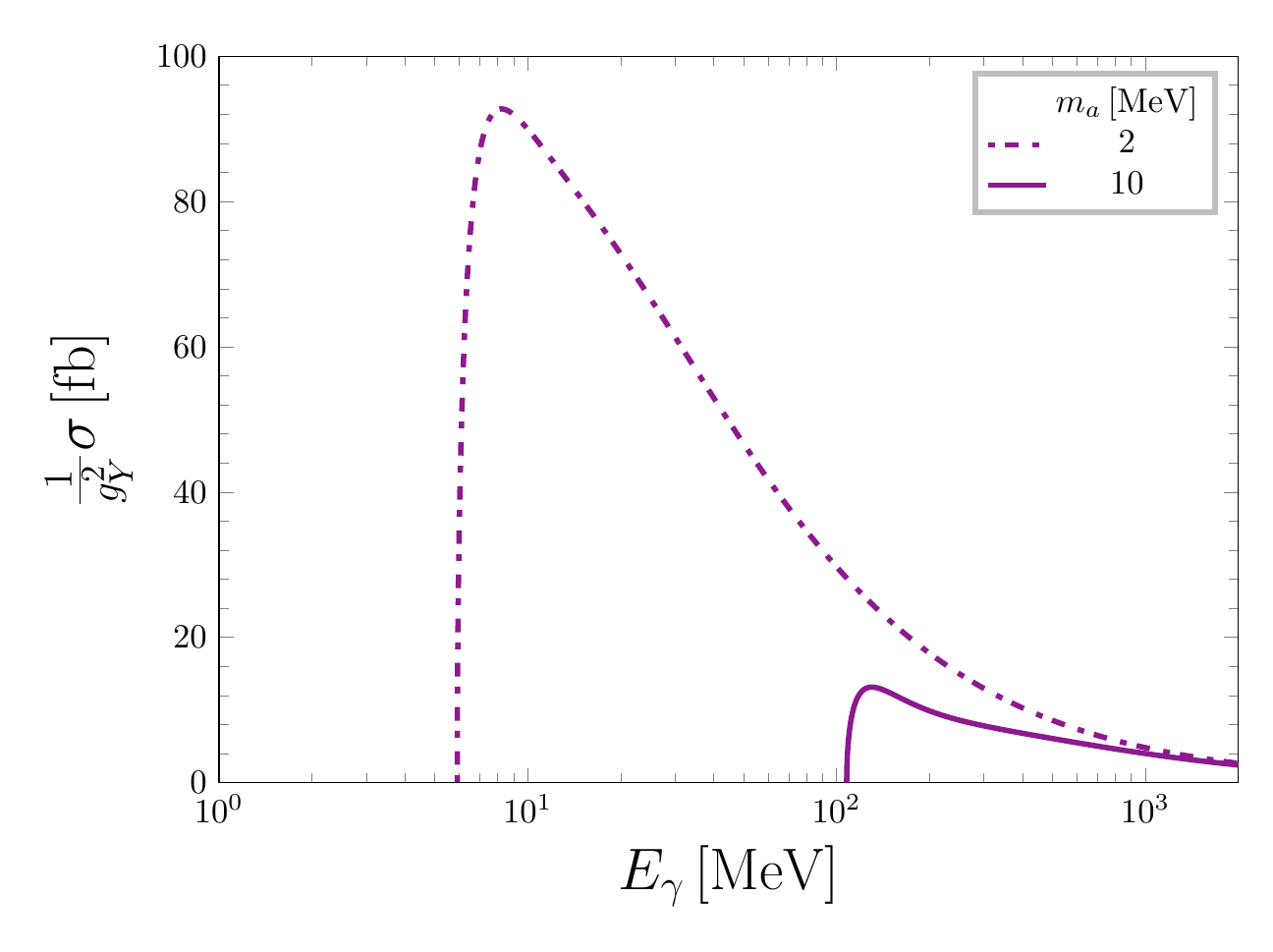}
  \caption{{ Production cross sections for dark Higgs bosons (left) and dark pseudoscalars (right) as a function of $E_{\gamma}$ for $m_{A'} = 2, 10~\mev$. } }
  \label{fig:sigscal}
\end{figure}

\begin{figure}[tb]
\begin{center}
\includegraphics*[width=0.49\textwidth]{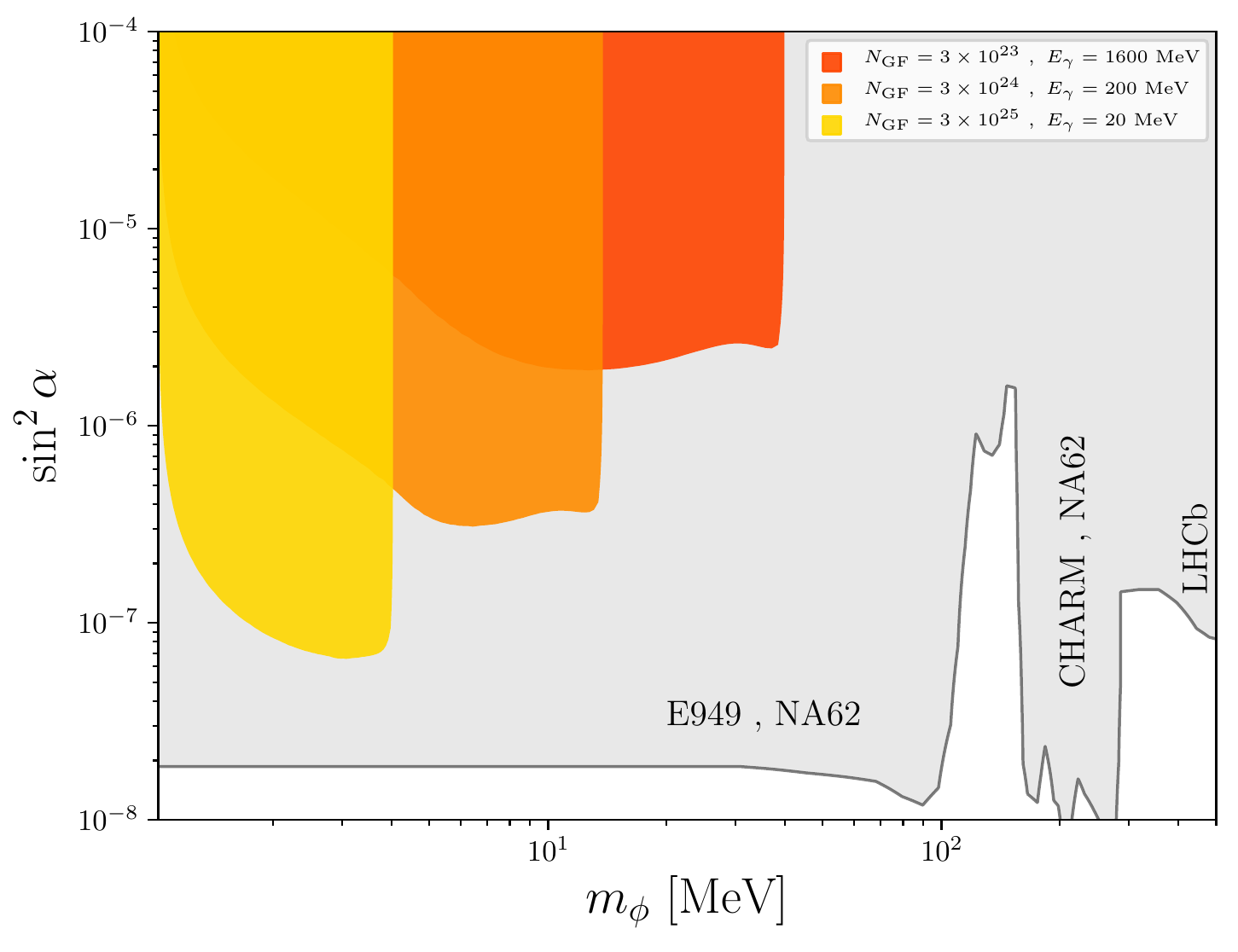}
\hfill
\includegraphics*[width=0.49\textwidth]{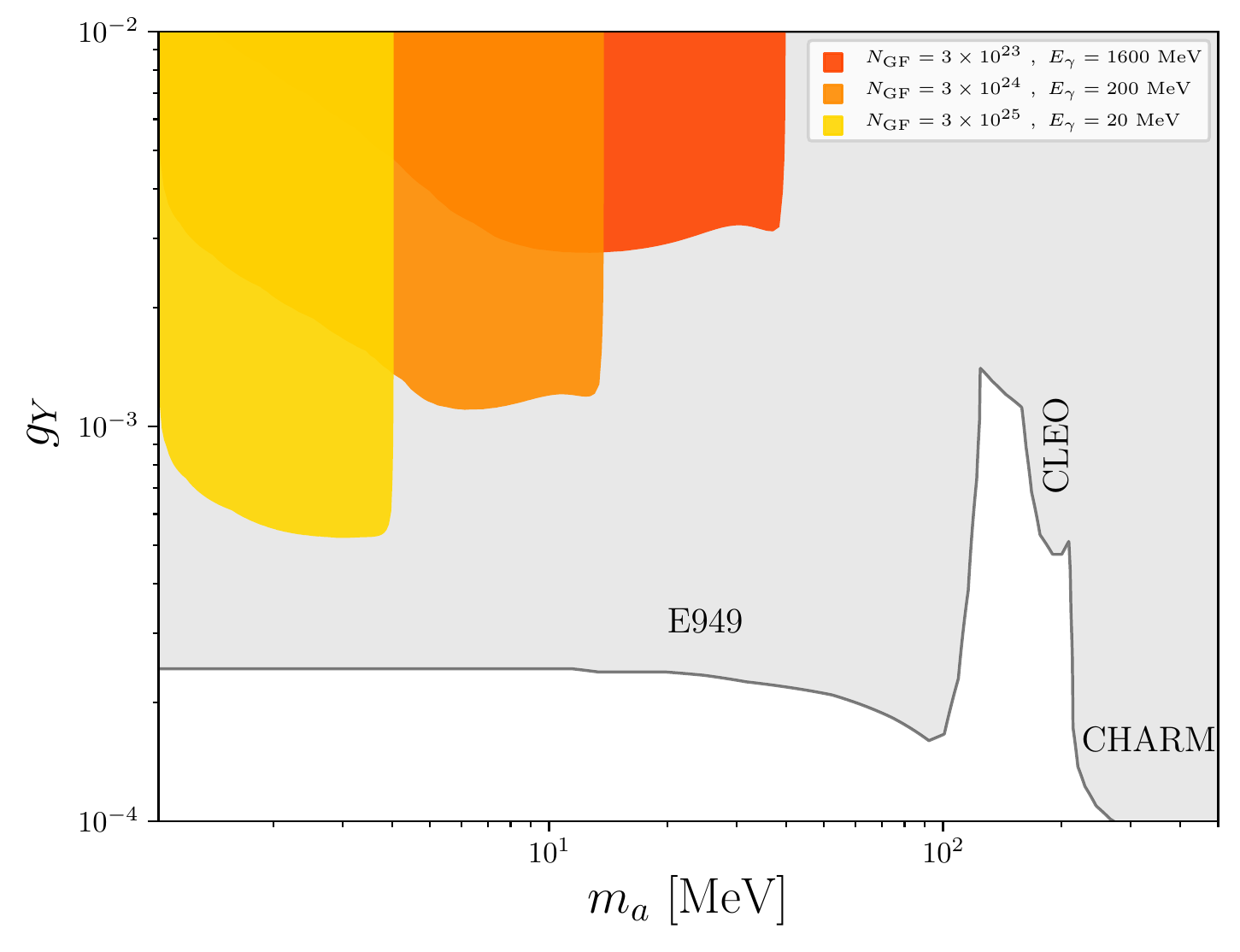}
\end{center}
\vspace*{-0.3in}
\caption{Sensitivity reaches for dark Higgs bosons (left) and dark pseudoscalars (right). The gray shaded regions indicate existing constraints from terrestrial experiments presented in Ref.~\cite{Beacham:2019nyx, CortinaGil:2021nts}.}
\label{fig:Spin0Reach}
\end{figure}

In \figref{Spin0Reach}, we show the GF sensitivity to these two spin-0 candidates. Unfortunately, the couplings of both spin-0 candidates considered here are Yukawa-suppressed.  This implies that the dark mediator's decays to electrons are extremely suppressed and the decay length is extremely long, which suppresses the rate.  Competing constraints, many of which use processes where the dark mediator interacts with a 2nd or 3rd generation particle and so is not as Yukawa-suppressed, are typically stronger, and the GF with one year of running does not probe new parameter space in these models.  

%%%%%%%%%%%%%%%%%%%%%%%%%%%%%%%%%%%%%%%%%%%
\section{Conclusions}
\label{sec:conclusions}
%%%%%%%%%%%%%%%%%%%%%%%%%%%%%%%%%%%%%%%%%%%

The proposed GF will be able to provide $10^{23}$ to $10^{25}$ photons on target per year, a remarkable leap in light source intensity.  By exploiting the LHC's ability to accelerate partially-stripped ions to Lorentz factors of $\gamma \sim 200 - 3000$, $\sim 10~\text{eV}$ photons can be back-scattered to 10 MeV to GeV energies, sufficient to search for new particles with masses in the $1-100$ MeV mass range.

In this paper, we have investigated for the first time the potential of the GF to discover new particles through dark Compton scattering, $\gamma e \to e X$, where $X$ is a dark photon, anomaly-free gauge boson, dark Higgs boson, or dark pseudoscalar.  In the cases of the spin-1 gauge bosons, we have found that the extraordinary intensities of the GF allow it to probe couplings as low as $\varepsilon \sim 10^{-9}$, over an order of magnitude lower than existing bounds from terrestrial experiments.  The $\varepsilon^4$ dependence of the signal event rate implies that as many as $10^4$ new gauge bosons may be produced in a year at the GF, or, in other words, the GF may start probing new models with just a few hours of running.  The region of parameter space with $\varepsilon \sim 10^{-9}$ can be probed by bounds from supernova cooling~\cite{Dent:2012mx,Dreiner:2013mua,Kazanas:2014mca,Rrapaj:2015wgs,Chang:2016ntp,Chang:2018rso,Sung:2019xie}, but such constraints depend on astrophysical assumptions that have been argued to weaken or possibly even remove them altogether~\cite{Bar:2019ifz}. The GF therefore provides a highly complementary probe.

The fixed target experiment proposed here is shown in \figref{experimentlayout}.  It consists of a low-$Z$ target to enhance the new physics event rate, followed by a high-$Z$ shield to eliminate SM background, followed by $\sim 10$ m-long decay volume and a tracking detector with a cross sectional area of $\sim 1-10~\m^2$.  We have assumed that the detection of coincident $e^+$ and $e^-$ particles that point back toward the GF photon beam, with an invariant mass equal to the $X$ boson's mass, will provide a spectacular and essentially background-free signal.

For the spin-0 candidates, with Yukawa-suppressed couplings to SM fermions, we have found poor discovery prospects, since the signal rates are highly suppressed by the GF's dependence on $X$ couplings to electrons.  For such models, GF photons scattering off not electrons, but nucleons and nuclei may provide significantly improved prospects.  Finally, we have considered only a small sample of the many possible new light, weakly-interacting particles. Axion-like particles have recently been considered~\cite{PBCMeetingBalkin}, and evaluations of the GF's sensitivity reaches for other particles, such as sterile neutrinos, may also be enlightening.

%%%%%%%%%%%%%%%%%%%%%%%%%%%%%%%%%%%
\acknowledgments
%%%%%%%%%%%%%%%%%%%%%%%%%%%%%%%%%%%

S.C.~acknowledges MHRD, Government of India, for a Senior Research Fellowship.  The work of J.L.F.~is supported in part by U.S.~National Science Foundation Grant No.~PHY-1915005 and by Simons Investigator Award \#376204. The work of M.V.~is supported in part by U.S.~National Science Foundation Grant No.~PHY-1915005.

\appendix

\section{Production Cross Section Calculations}

\begin{figure}[tb]
  \centering
  \includegraphics[width=0.8\textwidth]{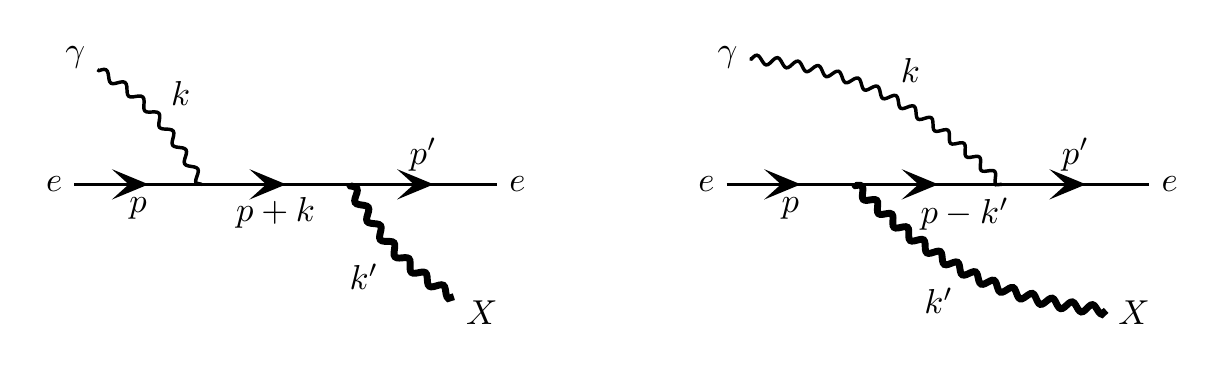}
  \caption{ Feynman diagrams contributing to the dark Compton scattering process $\gamma e \to e X$.}
  \label{fig:feyn_diag}
\end{figure}

In this Appendix, we derive the production cross sections entering the analysis.  The diagrams contributing to the ``dark Compton scattering'' processes $\gamma e \to e X$, where $X$ is a vector $A'$, a scalar $\phi$, or a pseudoscalar $a$, are shown in \figref{feyn_diag}.

Following the momentum assignments of \figref{feyn_diag}, the amplitude for the vector boson case is
\begin{align}
  {\cal M}_{A'}
  =&
  - g_X e
  \bar u (p^\prime) \left[ \frac{\slashed\epsilon^*_{k^\prime} (\slashed p + \slashed k + m_e) \slashed\epsilon_k}{s - m^2_e} + \frac{\slashed\epsilon_k (\slashed p - \slashed k^\prime + m_e) \slashed\epsilon^*_{k^\prime}}{u - m^2_e} \right] u (p) \ ,
\end{align}
where, for dark photons, $B-L$ gauge bosons, and $L_e - L_{\mu,\tau}$ gauge bosons, the coupling $g_X$ is $\varepsilon e$, $g_{B-L}$, and $g_{L_e - L_{\mu,\tau}}$, respectively. The spin-averaged amplitude squared is
\begin{align}
  \ol{|{\cal M}_{A'}|^2}
  =&
   g^2_X e^2
    \left\{
    4 (m^2_X + 2 m^2_e)\,m^2_e \left[ \frac{1}{s - m^2_e} + \frac{1}{u - m^2_e} \right]^2
    + 4 (m^2_X + 2 m^2_e) \left[ \frac{1}{s - m^2_e} + \frac{1}{u - m^2_e} \right]
    \right.
    \nl
    &\left.
    - 2 \left[ \frac{s - m^2_e}{u - m^2_e} + \frac{u - m^2_e}{s - m^2_e} \right]
    - 4 (m^2_X + 2 m^2_e) \frac{m^2_X}{(s - m^2_e) (u - m^2_e)}
    \right\}.
\end{align}
The amplitude squared has also been derived in Ref.~\cite{Chakrabarty:2019kdd}, and the above expression matches a similar expression found in Ref.~\cite{Liu:2017htz}, once one accounts for the different metric used. On integrating the differential cross section in the CM frame,
\begin{align}
  \frac{d\sigma^{\text{CM}}_{A'}}{d\cos\theta^*}
  =
  \frac{1}{32 \pi s} \frac{\lambda}{(s - m^2_e)}\,\ol{|{\cal M}_{A'}|^2} \ ,
  \label{eq:sigdiff_cm}
\end{align}
over the entire range of the angle $\theta^*$ between the incoming photon and the vector boson, one finds that the total cross section in the CM frame is
\begin{align}
  \sigma^{\text{CM}}_{A'}(s)
  =
  & \frac{ g^2_X e^2}{16 \pi}
  \left\{
  \lambda \left[ 8 \frac{(m^2_X + 2 m^2_e)}{(s - m^2_e)^3}
  + \frac{(s + m^2_e - m^2_X)}{s^2\,(s - m^2_e)} \right]
  \right.
  \nl
  &\left.
  + 2 \left[ \frac{1}{(s - m^2_e)} - 2 \frac{(m^2_X + 2 m^2_e)\,(s + m^2_e - m^2_X)}{(s - m^2_e)^3} \right] \ln\left|\frac{s + m^2_e - m^2_X + \lambda}{s + m^2_e - m^2_X - \lambda}\right|
  \right\},
  \label{eq:sigtot_cm}
\end{align}
where
\begin{align}
  \lambda = \sqrt{s^2 + m^4_X + m^4_e - 2s m^2_X - 2s m^2_e - 2 m^2_X m^2_e} \ .
\end{align}

In the lab frame, where the photon is scattered off a static electron, the differential cross section can be obtained from the expression of \eqref{sigdiff_cm} by applying a Lorentz boost along the opposite direction to the incoming electron in the CM frame to bring it to rest. Therefore, in the lab frame, the differential cross section for vector boson production will be
\begin{align}
 \frac{d\,\sigma^{\text{lab}}_{A'}}{d\cos\theta}
 =
 \frac{d\,\sigma^{\text{CM}}_{A'}}{d\cos\theta^*}
 \frac{d\cos\theta^*}{d\cos\theta} \ ,
 \quad\text{using}\quad
 \cos\theta^*
 =
 \frac{\gamma\,( \cos\theta - \beta/\beta_{A'} )}{\sqrt{\sin^2\theta + \gamma^2\,( \cos\theta - \beta/\beta_{A'} )^2}} \ ,
\end{align}
where $\beta$ and $\beta_{A'}$ are, respectively, the velocity of the lab frame with respect to the CM frame and the velocity of the scattered vector boson along the direction of its scattering angle $\theta$ in the lab frame. As usual, $\gamma = 1/\sqrt{1 - \beta^2}$.

In principle, the total cross section in the lab frame can be derived by integrating the above differential cross section over the entire range of the scattering angle $\theta$. However, for a massive vector boson, the integration can be non-trivial. On the other hand, since the total cross section is boost-invariant, we can safely bypass the intricacies of such integration by simply substituting 
\begin{align}
  s &= (p + k)^2 = m^2_e + 2 m_e\,E_\gamma  
  \label{eq:s_rest} \\
  \lambda\,(E_\gamma) &= \sqrt{(2 m_e\,E_\gamma - m^2_X)^2 - 4 m^2_X m^2_e} 
  \label{eq:lambda_rest}
\end{align}
in \eqref{sigtot_cm} to find that the total cross section in the lab frame is
\begin{align}
  \sigma^{\text{lab}}_{A'}(E_\gamma)
  &=
  \frac{g^2_X e^2}{16 \pi}
  \left\{
  \lambda (E_\gamma) \left[ \frac{(m^2_X + 2 m^2_e)}{m^3_e\,E_\gamma^3}
  + \frac{(2 m_e\,E_\gamma + 2 m^2_e - m^2_X)}{2 m_e\,E_\gamma\,(2 m_e\,E_\gamma + m^2_e)^2} \right]
  \right.
  \nl
  &\hskip-3em\left.
  + \left[ \frac{1}{m_e E_\gamma} \! - \! \frac{(m^2_X \! + \! 2 m^2_e)\,(2 m_e E_\gamma \! + \! 2 m^2_e \! - \! m^2_X)}{2 m^3_e\,E_\gamma^3} \right] \ln\left|\frac{2 m_e E_\gamma + 2 m^2_e - m^2_X + \lambda (E_\gamma)}{2 m_e E_\gamma + 2 m^2_e - m^2_X - \lambda (E_\gamma)}\right|
  \right\} ,
  \label{eq:sigtot_lab}
\end{align}
where $E_\gamma$ is the energy of the incident photon. From \eqref{s_rest}, we can also find that the threshold photon energy for $X$ production is
\begin{align}
  E_{\gamma}^{\text{th}}
  =
  m_X + \frac{m^2_X}{2 m_e} \ .
\end{align}

In a similar way as above, we can derive the corresponding expressions for the dark Higgs boson and dark pseudoscalar cases. The corresponding amplitudes are
\begin{align}
  {\cal M}_\phi
  =&
  - g_X e \,
  \bar u (p^\prime) \left[ \frac{(\slashed p + \slashed k + m_e) \slashed\epsilon_k}{s - m^2_e} + \frac{\slashed\epsilon_k (\slashed p - \slashed k^\prime + m_e)}{u - m^2_e} \right] u (p) \ ,
  \\
  {\cal M}_a
  =&
  - g_X e \,
  \bar u (p^\prime) \left[ \frac{\gamma_5 (\slashed p + \slashed k + m_e) \slashed\epsilon_k}{s - m^2_e} + \frac{\slashed\epsilon_k (\slashed p - \slashed k^\prime + m_e) \gamma_5}{u - m^2_e} \right] u (p) \ ,
\end{align}
where, for dark Higgs bosons and dark pseudoscalars, the coupling $g_X$ is $\sin\alpha\,m_e/v$ and  $g_Y\,m_e/(2v)$, respectively. 

The spin-averaged matrix elements squared are the same as in Refs.~\cite{Liu:2016mqv,Liu:2017htz} (with the appropriate choice of metric):
\begin{align}
  \ol{|{\cal M}_\phi|^2}
  &=
  g^2_X e^2
  \left\{
  2 (m^2_X - 4 m^2_e) m^2_e \left[ \frac{1}{s - m^2_e} + \frac{1}{u - m^2_e} \right]^2
  + 2 (m^2_X - 4 m^2_e) \left[ \frac{1}{s - m^2_e} + \frac{1}{u - m^2_e} \right]
  \right.
  \nl
  &\left.
  - \left[ 2 + \frac{s - m^2_e}{u - m^2_e} + \frac{u - m^2_e}{s - m^2_e} \right]
  - 2 (m^2_X - 4 m^2_e) \frac{m^2_X}{(s - m^2_e) (u - m^2_e)}
  \right\},
  \\
  \ol{|{\cal M}_a|^2}
  &=
  g^2_X e^2
  \left\{
  2 m^2_X m^2_e \left[ \frac{1}{s - m^2_e} + \frac{1}{u - m^2_e} \right]^2
  + 2 m^2_X \left[ \frac{1}{s - m^2_e} + \frac{1}{u - m^2_e} \right]
  \right.
  \nl
  &\left.
  - \left[ 2 + \frac{s - m^2_e}{u - m^2_e} + \frac{u - m^2_e}{s - m^2_e} \right]
  - \frac{2 m^4_X}{(s - m^2_e) (u - m^2_e)}
  \right\}.
\end{align}
The resulting expressions for the total cross sections in the CM frame are
\begin{align}
  \sigma^{\text{CM}}_\phi (s)
  &=
  \frac{g^2_X e^2}{16 \pi}
  \left\{
  \lambda\,\left[ 4 \frac{(m^2_X - 4 m^2_e)}{(s - m^2_e)^3}
  + \frac{(m^2_e - 3s-m^2_X)}{2s^2\,(s - m^2_e)} \right]
  \right.
  \nl
  &\left.
  + \left[ \frac{1}{(s - m^2_e)} - 2 \frac{(m^2_X - 4 m^2_e)\,(s + m^2_e - m^2_X)}{(s - m^2_e)^3} \right] \ln\left|\frac{s + m^2_e - m^2_X + \lambda}{s + m^2_e - m^2_X - \lambda}\right|
  \right\},
  \\
  \sigma^{\text{CM}}_a (s)
  &=
  \frac{g^2_X e^2}{16 \pi}
  \left\{
  \lambda\,\left[ 4 \frac{m^2_X}{(s - m^2_e)^3}
  + \frac{(m^2_e - 3s-m^2_X)}{2s^2\,(s - m^2_e)} \right]
  \right.
  \nl
  &\left.
  + \left[ \frac{1}{(s - m^2_e)} - 2 \frac{m^2_X \,(s + m^2_e - m^2_X)}{(s - m^2_e)^3} \right] \ln\left|\frac{s + m^2_e - m^2_X + \lambda}{s + m^2_e - m^2_X - \lambda}\right|
  \right\}.
\end{align}
Finally, the expressions for the total cross sections in the lab frame are
\begin{align}
  \sigma^{\text{lab}}_\phi (E_\gamma)
  &=
  \frac{g^2_X e^2}{16 \pi}
  \left\{
  \lambda (E_\gamma) \left[ \frac{(m^2_X - 4 m^2_e)}{2m^3_e\,E_\gamma^3}
  - \frac{(6 m_e\,E_\gamma + 2 m^2_e + m^2_X)}{4 m_e\,E_\gamma\,(2 m_e\,E_\gamma + m^2_e)^2} \right]
  \right.
  \nl
  &\hskip-3em\left.
  + \left[ \frac{1}{2m_e E_\gamma} \! - \! \frac{(m^2_X \! - \! 4 m^2_e) (2 m_e E_\gamma \! + \! 2 m^2_e \! - \! m^2_X)}{4 m^3_e\,E_\gamma^3} \right] \ln\left|\frac{2 m_e E_\gamma \! + \! 2 m^2_e \! - \! m^2_X \! + \! \lambda(E_\gamma)}{2 m_e E_\gamma \! + \! 2 m^2_e \! - \! m^2_X \! - \! \lambda(E_\gamma)}\right|
  \right\} \! ,
  \\
  \sigma^{\text{lab}}_a (E_\gamma)
  &=
  \frac{ g^2_X e^2}{16 \pi}
  \left\{
  \lambda\,(E_\gamma)\,\left[ \frac{m^2_X}{2m^3_e\,E_\gamma^3}
  - \frac{(6 m_e\,E_\gamma + 2 m^2_e + m^2_X)}{4 m_e\,E_\gamma\,(2 m_e\,E_\gamma + m^2_e)^2} \right]
  \right.
  \nl
  &\hskip-3em\left.
  + \left[ \frac{1}{2m_e\,E_\gamma} - \frac{m^2_X\,(2 m_e\,E_\gamma + 2 m^2_e - m^2_X)}{4 m^3_e\,E_\gamma^3} \right] \ln\left|\frac{2 m_e\,E_\gamma + 2 m^2_e - m^2_X + \lambda\,(E_\gamma)}{2 m_e\,E_\gamma + 2 m^2_e - m^2_X - \lambda\,(E_\gamma)}\right|
  \right\}.
\end{align}

\bibliography{gfbsmbib}

\end{document}